\documentclass[10pt,twocolumn,twoside]{IEEEtran}
\usepackage{latexsym}
\usepackage{amsmath}
\usepackage{amssymb}
\usepackage{epsfig}
\usepackage{cite}
\usepackage{algorithmic}
\usepackage{overpic}
\usepackage{multirow}
\usepackage[table]{xcolor}
\usepackage{graphicx}
\usepackage{colortbl,array}
\usepackage{multirow,bigdelim}
\usepackage{graphicx}
\usepackage{subfigure}
\usepackage{booktabs}
\usepackage{booktabs,amsmath}
\usepackage{fancyhdr, graphicx, amsmath, amssymb}
\usepackage{nomencl}
\newtheorem{thm}{Theorem}
\newtheorem{lem}{Lemma}
\newtheorem{rem}{Remark}
\newtheorem{cor}{Corollary}
\newtheorem{asm}{Assumption}
\newtheorem{defa}{Definition}

\ifCLASSINFOpdf
\else
\fi

\begin{document}

\title{Impact of Confirmation Bias on Competitive Information Spread in Social Networks}

\author{Yanbing~Mao, Emrah~Akyol, and Naira~Hovakimyan
\thanks{Y.~Mao and N.~Hovakimyan are with the Department of Mechanical Science and Engineering, University of Illinois at Urbana--Champaign, Urbana, IL, 61801 USA (e-mail: \{ybmao, nhovakim\}@illinois.edu).}
\thanks{E.~Akyol is with the Department of Electrical and Computer Engineering, Binghamton University--SUNY, Binghamton, NY,
13902 USA (e-mail: eakyol@binghamton.edu). }
\thanks{Parts of the material in this paper were presented in the 53rd Annual Asilomar Conference on Signals, Systems, and Computers, 2019 \cite{asilomar}.}
}

\maketitle

\begin{abstract}
This paper investigates the impact of confirmation bias on competitive information spread in the social network that comprises individuals in a social network and competitive information sources. We formulate the problem as a zero-sum game, which admits a unique Nash equilibrium in pure strategies. We characterize the dependence of pure Nash equilibrium on the public's innate opinions, the social network topology, as well as the parameters of confirmation bias. We uncover that confirmation bias moves the equilibrium towards the center only when the innate opinions are not neutral, and this move does not occur for the competitive information sources simultaneously. Numerical examples in the context of well-known Krackhardt's advice network are provided to demonstrate the correctness of theoretical results.
\end{abstract}
\IEEEpeerreviewmaketitle

\section{Introduction}
Mathematical models for the opinion formation in networks
has been an important research subject for decades, see e.g., \cite{moreno1934shall,french1956formal}. A few well-known models include DeGroot model \cite{degroot1974reaching} (whose roots go back to \cite{french1956formal, abelson1964mathematical}) that considers opinion evolution within a network in terms of the weighted average of individuals' connections where weights are determined by influences.  Friedkin-Johnsen model~\cite{friedkin1990social} incorporates individual innate opinions, thereby making the model more suitable to several real-life scenarios, as well as real applications, e.g., optimal investment for competing camps \cite{dhamal2018optimal} and debiasing social influence \cite{das2013debiasing}. In  \cite{hegselmann2002opinion}, a bounded confidence model is presented where individuals are influenced by their neighbors that are not too far from their opinion, modeling the main subject of the paper that is confirmation bias (CB). Majority of the recent works are variations of these models, with few exceptions.  An overview of opinion dynamic models can be found in relevant tutorial papers, see e.g., \cite{proskurnikov2017tutorial,proskurnikov2018tutorial}, and the references therein, for an excellent overview.

While opinion evolution models have always been an active research area, recently with the wide use of social media \cite{xu2020paradox}, in conjunction with  automated news  generation with the help of artificial intelligence technologies \cite{giridhar2019social,cui2019semi}, it has gained a vital importance in studying misinformation spread and polarization. In this regard, CB plays a key role. CB broadly refers to cognitive bias towards favoring information sources that affirm existing opinion \cite{nickerson1998confirmation}. It is well understood that CB helps create ``echo chambers" within networks, in which misinformation and polarization thrive, see e.g., \cite{lazer2018science, vicario2019polarization,kappes2020confirmation}.

In this paper, we study competitive information spread in social networks, with a particular focus on the impact of CB on the results. Competitive information spread has been studied extensively in recent years. In the following, we review a few relevant recent studies. \textcolor[rgb]{0.00,0.00,1.00}{Building on DeGroot model \cite{degroot1974reaching}, Zhao et al. in \cite{zhao2016competitiveness} investigated how to enhance a competitor's competitiveness through adding new communication links to normal agents to maximize the number of supporters or the supporting degree towards a competitor. Rusinowska and Taalaibekova in \cite{rusinowska2019opinion} proposed
a model of competitive opinion formation with three persuaders, who respectively hold extreme, opposite and centrist opinions, while Grabisch et al. in \cite{grabisch2018strategic} investigated the model of influence with a set of nonstrategic agents and two
strategic agents that have fixed but opposed opinions.} Dhamal et al. in~\cite{dhamal2018optimal} incorporated opponent stubborn agents into Friedkin-Johnsen model \cite{friedkin1990social}. Employing a diffusion dynamics, Eshghi et al. in~\cite{eshghi2018spread} studied optimal allocating of a finite budget across several advertising channels. \textcolor[rgb]{0.00,0.00,1.00}{Meanwhile, Proskurnikov et al. in~\cite{proskurnikov2016opinion} studied the opinion dynamics with negative weights, which model antagonistic or competitive interactions, with its origins
dating back to the seminal work of Altafini \cite{altafini2012consensus}}. We note however that these prior works on competitive camps and competitive/antagonistic interactions do not consider CB in their analysis.

Among the aforementioned opinion evolution models, CB can be modeled within the context of bounded confidence models such as the Hegselmann-Krause model \cite{hegselmann2002opinion} and its recent variations \cite{yang2014opinion,pineda2013noisy}. However, these models involve a discontinuity in the influence impact: an individual is either influenced by an information source (or her neighbors) fully or not at all, depending on the opinion differences. This binary influence effect renders the analysis of the steady-state point difficult in general. As a remedy, in \cite{mao2018spread}, a new opinion dynamics model is proposed as a variation of Friedkin-Johnsen model \cite{friedkin1990social} with continuous bias model.

In this paper, building on preliminary analysis in \cite{asilomar}, we analyze the information spread over a network with two competitive information sources, where the only control variables are the opinions of information sources. We adopt the opinion dynamics in \cite{mao2018spread} with two information sources and a (state-dependent) piecewise linear CB model. We formulate the problem as a zero-sum game and show that this game admits a unique Nash equilibrium which is in pure strategies.  We particularly study the impact of CB on the Nash equilibrium. We analyze how the equilibrium achieving strategies depend on public's innate opinions, the network topology, as well as the CB parameters.

This paper is organized as follows. In Section II, we present preliminaries. The problem is formulated in Section III. In Section IV, we investigate the Nash equilibrium. We study the impact of CB and innate opinions on Nash equilibrium in Section V. We present numerical simulations in Section VI. We finally present our conclusions and future research directions in Section VII.

\vspace{-0.00cm}
\section{Preliminaries}
\subsection{Notation}
\label{notation}
Let $\mathbb{R}^{n}$ and $\mathbb{R}^{m \times n}$ denote the set of $\emph{n}$-dimensional real vectors and the set of $m \times n$-dimensional real matrices, respectively. $\mathbb{N}$ represents the set of the positive integers, and $\mathbb{N}_{0} = \mathbb{N} \cup \{0\}$. We define $I$ as the identity  matrix with proper dimension. We let $\mathbf{1}$ denote the vector of all ones. The superscript `$\top$' stands for the matrix transposition. For a vector $x \in \mathbb{R}^{n}$, $\left\| x \right\|$ stands for its $l_{1}$ norm, i.e., $\left\| x \right\| = \sum\limits_{i = 1}^n {\left| {{x_i}} \right|}$. For $W$ $=$ $\left[ {{w_{ij}}} \right] \in {\mathbb{R}^{n \times n}}$, we use
${\left\| W \right\|_1}$ and ${\left\| W \right\|_\infty }$ to denote $\mathop {\max }\limits_{j = 1, \ldots,n} \{\sum\limits_{i = 1}^n {\left| {{w_{ij}}} \right|} \}$ and $\mathop {\max }\limits_{i = 1, \ldots ,n}\{\sum\limits_{j = 1}^n {\left| {{w_{ij}}} \right|}\}$, respectively.

The social network considered in this paper is composed of $n$ individuals. The interaction among the individuals is modeled by a digraph $\mathfrak{G} = (\mathbb{V}, \mathbb{E})$, where $\mathbb{V}$ = $\left\{\mathrm{v}_{1}, \ldots,  \mathrm{v}_{\mathrm{n}}\right\}$ is a set of vertices representing the individuals and $\mathbb{E} \subseteq  \mathbb{V} \times \mathbb{V}$ is a set of edges representing the influence structure. We take the network to have no self-loops, i.e., for any $\mathrm{v}_{\mathrm{i}} \in \mathbb{V}$, we assume that $(\mathrm{v}_{\mathrm{i}}, \mathrm{v}_{\mathrm{i}}) \notin \mathbb{E}$.

\vspace{-0.00cm}
\subsection{Opinion Dynamics}
In this paper, we adopt the opinion evolution model in the presence of two competitive information sources in \cite{mao2018spread}:
\begin{align}
&{x_i}(k + 1) = {\alpha _i}(x_{i}(k)){s_i} + \sum\limits_{j \in \mathbb{V}}{{w_{ij}}{x_j}(k)}  + {{\overline{w}}(x_{i}(k)){h}} \nonumber \\
&\hspace{3.55cm}+ {{{\underline{w}}}(x_{i}(k)){g}},~~i \in \mathbb{V}.  \label{eq:od}
\end{align}
In the following, we describe the elements of this model:
\begin{enumerate}
  \item $x_i(k) \in [0,1]$ is individual $\mathrm{v}_{\mathrm{i}}$'s opinion at time $k$. This opinion evolves in time as described in (\ref{eq:od}).
  \item $s_i \in [0,1]$ is  individual $\mathrm{v}_{\mathrm{i}}$'s  innate opinion which is fixed in time. We define the extremal innate opinions as follows:
\begin{align}
\overline{s} \triangleq  \mathop {\max }\limits_{i \in \mathbb{V}} \left\{ {{s_i}} \right\}, ~~~~~\underline{s} \triangleq \mathop {\min }\limits_{i \in \mathbb{V}} \left\{ {{s_i}} \right\}.\nonumber
\end{align}

  \item $w_{ij}$ represents the influence of individual $\mathrm{v}_{\mathrm{j}}$ on  $\mathrm{v}_{\mathrm{i}}$,
  \begin{equation}
	w_{ij} = \begin{cases}
		> 0, & \text{if } (\mathrm{v}_\mathrm{i},\mathrm{v}_\mathrm{j}) \in \mathbb{E}\\
		= 0, & \text{otherwise}.
	\end{cases} \nonumber
  \end{equation}
 This is a standard model parameter, with its origins dating back to the seminal work of \textcolor[rgb]{0.00,0.00,1.00}{Friedkin and Johnsen} \cite{friedkin1990social}. In this paper, we do not consider antagonistic interactions, as studied in \cite{proskurnikov2016opinion}, which would imply negative values for $w_{ij}$.   
  \item $h$ and $g$ are the opinions of competitive information sources (or stubborn individuals), \textcolor[rgb]{0.00,0.00,1.00}{Hank and Georgia}, respectively. Their objectives are to move the public opinion to two extremes they represent. We assume that the values of $g,h$ satisfy the following:
      \begin{align}
 1 \geq h \geq \overline{s} \geq \underline{s} \geq g \geq 0
\label{eq:oofn}
\end{align}
This assumption states that the information sources are more extremal than the most extreme innate opinion of the public, prior to any external influence.
  \item $\overline{w}(x_{i}(k))$ and $\underline{w}(x_{i}(k))$ are the state-dependent influence weights of information sources \textcolor[rgb]{0.00,0.00,1.00}{Hank and Georgia} on individual $\mathrm{v}_{\mathrm{i}}$. These weights model confirmation bias as
   \begin{subequations}
  \begin{align}
  \overline{w}(x_{i}(k)) = \beta - \gamma \left| x_i(k) - h \right|, \label{eq: sdw1}\\
  \underline{w}(x_{i}(k)) = \beta - \gamma \left| x_i(k) - g \right|, \label{eq: sdw2}
  \end{align}\label{eq:sdw}
  \end{subequations}
  \!\!\!where $\beta \in \mathbb R$ and  $\gamma \in \mathbb R$ are bias parameters.  Throughout this paper, we make the following assumption on the bias parameters and influence weights:
  \begin{asm} Given $W \in \mathbb{R}^{n \times n}$, $\beta \in \mathbb{R}$ and $\gamma \in \mathbb{R}$: \begin{subequations}
\begin{align}
&\beta  \geq \gamma \geq 0,\label{eq:asfx00}\\
&1 - \max \left\{ {{{\left\| W \right\|}_\infty },{{\left\| W \right\|}_1}} \right\} \ge \max \left\{ {2\beta ,4\gamma } \right\}, \label{eq:asfx}\\
&\textcolor[rgb]{0.00,0.00,1.00}{W~\text{has a positive eigenvector}} \label{adjpo}.
\end{align}\label{eq:hhk}
\end{subequations}
\label{assum1}
\end{asm}
\vspace{-5.5mm}
Here, the CB model (3) is assumed to be piecewise linear. We note that in the original model used in  \cite{mao2018spread}, this bias function is taken in general possibly nonlinear and decreasing.
  \item $\alpha_i(x_{i}(k))$  is the ``resistance parameter'' of individual $\mathrm{v}_{\mathrm{i}}$ and is determined in such a way that it satisfies
      \begin{align}
      \!\!\!{\alpha _i}(x_{i}(k)) + \sum\limits_{j \in \mathbb{V}}\! {{w_{ij}} +  \overline{w}( x_{i}(k)  ) +  \underline{w}(x_{i}(k))} = 1,\label{eq:as122f}
      \end{align}
      $\forall i \in \mathbb{V}$ and $\forall k \in \mathbb{N}_{0}$. This is a standard assumption common in all classical opinion dynamics models, see e.g., \cite{friedkin1990social,das2013debiasing,dhamal2018optimal}. The entire model essentially represents that individuals form opinions by taking weighted averages over a convex polytope of different contributing factors.
\end{enumerate}

\vspace{-0.20cm}
\textcolor[rgb]{0.00,0.00,1.00}{\begin{rem}
Confirmation bias refers to the tendency to acquire or process new information in a way that confirms one's preconceptions and avoids contradiction with prior belief \cite{nickerson1998confirmation}. We note that function \eqref{eq:sdw} is more like state-dependent social influence weights used to model homophily \cite{mas2014cultural, duggins2014psychologically}, which is used in this paper to describe the confirmation bias behavior to some extent. This is motivated by following observations:
\begin{itemize}
  \item Both polarization and homogeneity are the results of the conjugate effect of confirmation bias and social influence \cite{del2017modeling,del2016spreading}.
  \item Confirmation bias happens when a person gives more weight to evidence that confirms their beliefs and undervalues evidence that could disprove it.
\end{itemize}
\end{rem}
}

\vspace{-0.60cm}
\textcolor[rgb]{0.00,0.00,1.00}{\begin{rem}
We obtain from \eqref{eq:sdw} and \eqref{eq:as122f} that
\begin{align}
{\alpha _i}(x_{i}(k))&= 1-\sum\limits_{j \in \mathbb{V}} \!{{w_{ij}} -  \overline{w}( x_{i}(k)  ) -  \underline{w}(x_{i}(k))} \nonumber\\
&=1-\!\sum\limits_{j \in\mathbb{V}}\!{w_{ij}}\!-\!2\beta\!+\!\gamma\left({\left| {{x_i}(k)\!-\!h}\right|+\!\left|{{x_i}(k)\!-\!g}\right|}\right)\nonumber \\
& \ge 1 - \sum\limits_{j \in \mathbb{V}}  {w_{ij}} - 2\beta. \label{nn2}
\end{align}
The inequality \eqref{nn2} indicates that to guarantee the non-negativeness of ${\alpha _i}(x_{i}(k))$, we require $1 - \sum\limits_{j \in \mathbb{V}}  {w_{ij}} \geq 2\beta$ for any $i \in \mathbb{V}$, or equivalently,
\begin{align}
1 - {\left\| W \right\|_\infty } \ge 2\beta, \label{com1}
\end{align}
which holds under Assumption \ref{assum1}. We note that Assumption \ref{assum1}, in conjunction with \eqref{eq:as122f}, also guarantees the non-negativeness of state-dependent influence weights \eqref{eq:sdw} and the convergence of dynamics in \eqref{eq:od}, whose detailed proofs are included in the proof of Theorem \ref{thm:th1} (see Appendix B).
\end{rem}
}

\vspace{-0.30cm}
\textcolor[rgb]{0.00,0.00,1.00}{\begin{rem}
We assume that influence weights $w_{ij}$, $i,j \in \mathbb{V}$, are fixed since the social influence among individuals is based on ``trust," which
tends to vary little over a long period of time. However, the influence of information sources over individuals depends heavily on the current opinions of individuals, due to the
confirmation bias. This is why the weight of influence of information source on individual defined as \eqref{eq:sdw} is state-dependent.
\end{rem}
}

We next express (\ref{eq:od}) in the  vector form:
\begin{align}
\!\!{x}({k \!+\! 1}) \!=\! \mathcal{A}(x(k))s \!+\! Wx(k) \!+\! \overline{\mathcal{W}}(x(k))h \!+\! \underline{\mathcal{W}}(x(k))g,
\label{eq:odd}
\end{align}
where we define:
\begin{subequations}
\begin{align}
s & \triangleq {\left[ {{s_{1}}, \ldots ,{s_{n}}} \right]^\top} \in \mathbb{R}^{n},\\
x(k) & \triangleq {\left[ {{x_{1}}(k), \ldots ,{x_{n}}(k)} \right]^ \top} \in \mathbb{R}^{n},\\
W & \triangleq \left[ {{w_{ij}}} \right] \in {\mathbb{R}^{n \times n}},\\
\mathcal{A}(x(k)) &\triangleq \text{diag}\!\left\{ {{\alpha _1}({x_1}(k)), \ldots, {\alpha _n}({x_n}( k ))} \right\} \!\in\! {\mathbb{R}^{n \times n}}\!,\\
\overline{\mathcal{W}}( {x( k )}) &\triangleq {\left[ {{{\overline w}}( {{x_1}( k )}), \ldots ,{{\overline w}}( {{x_n}( k )})} \right]^\top} \in {\mathbb{R}^n},\\
\underline{\mathcal{W}}({x(k)}) &\triangleq {\left[ {{{\underline w}}( {{x_1}( k )}), \ldots ,{{\underline w}}( {{x_n}( k )})} \right]^\top} \in {\mathbb{R}^n}.
\end{align}
\end{subequations}

\textcolor[rgb]{0.00,0.00,1.00}{In \cite{mao2018spread},  it is shown that similar dynamics converge to a unique steady-state, independent of the initial opinions, for more general bias functions and information sources. Here, we show that the derived convergence condition \eqref{eq:hhk} is more relaxed for this more specific model.} Moreover, we analytically analyze the steady-state point achieved  by the opinion dynamics. \textcolor[rgb]{0.00,0.00,1.00}{We reiterate that the primary advantage of the model described in \eqref{eq:od}, in contrast with the classical bounded confidence models such as the Hegselmann-Krause model \cite{hegselmann2002opinion}, is that \eqref{eq:od} allows us to examine the steady-state point analytically. This is because the state-dependent weights in the classical models can be equal to zero when the opinion distance is larger than the confidence bound, which renders analysis difficult, while state-dependent weights in \eqref{eq:od} are nonzero for almost all scenarios. For an analytical expression, similar settings are imposed on state-dependent susceptibility of polar opinion dynamics \cite{amelkin2017polar,liu2018discrete}}. Before stating our formally, we  define the following  matrices:
\begin{align}
E &\triangleq I - W + \left( {g - h} \right)\gamma I \in \mathbb{R}^{n \times n},  \label{eq:def1}\\
D &\triangleq \text{diag}\left\{ {\sum\limits_{j \in \mathbb{V}} \!w_{1j}, \sum\limits_{j \in \mathbb{V}} \!w_{2j}, \ldots, \sum\limits_{j \in \mathbb{V}} \!w_{nj}} \right\} \in \mathbb{R}^{n \times n}.
\end{align}

With these definitions at hand, we present our convergence result, whose proof appears in Appendix B.
\begin{thm}
For any  $\mathbf x(0)$, the dynamics in \eqref{eq:od} converge to
\begin{align}
x^*(g,h) & = {E^{ - 1}}\left( {{I} - D - 2\beta {I} + ({h - g})\gamma I} \right)s \nonumber\\
&\hspace{1.3cm} + {E^{ - 1}}\left( {(h + g)\beta {{\bf{1}}} + ({g^2} - {h^2})\gamma {{\bf{1}}}} \right).
\label{eq:mmeq}
\end{align}\label{thm:th1}
\end{thm}
\vspace{-0.90cm}
\textcolor[rgb]{0.00,0.00,1.00}{\begin{rem}
In light of Gershgorin circle theorem, we straightforwardly verify from \eqref{eq:hhk} and \eqref{eq:def1} that all of the eigenvalues of $E$ are nonzero, and thus, $E$ is invertible.
\end{rem}}

\section{Problem Formulation}
In this work, we analyze the values of information sources (or stubborn individuals as referred in some prior work, e.g., \cite{dhamal2018optimal}) \textcolor[rgb]{0.00,0.00,1.00}{Hank and Georgia} would provide in a setting, where they strive to move the steady-state opinion (whose exact expression is provided in Theorem \ref{thm:th1}) of the network to the two binary extremes. We note that while prior work \cite{dhamal2018optimal} has studied similar problems, albeit without CB, with respect to resource allocation and node placement within a network as the decision variable, here we use the opinion values $g$ and $h$ (\textcolor[rgb]{0.00,0.00,1.00}{Hank and Georgia} provide to the network) as the decision variables, and keep the topology of social network constant. This problem constitutes an unconstrained zero-sum game between \textcolor[rgb]{0.00,0.00,1.00}{Hank and Georgia}, with continuous strategy  spaces $g \in [0,\underline{s}]$ and $h \in [\overline{s},1]$, for which Nash equilibria are sought.

At first glance, it might be tempting to conclude that the trivial choice of \textcolor[rgb]{0.00,0.00,1.00}{$g=0$ and  $h=1$} are the equilibrium achieving strategies for \textcolor[rgb]{0.00,0.00,1.00}{Hank and Georgia}. Indeed, we formally show that (in Sections \ref{nocbba} and \ref{nocb}, Corollaries \ref{corfn} and \ref{nercc}) these strategies are equilibrium-achieving, in the absence of CB, or in the special case of neutral innate opinions (i.e., $s_{i} = \frac{1}{2}, \forall i \in \mathbb{V}$) even in the presence of CB.

However, this is exactly the aspect in which CB renders this problem a formidable research challenge. The strategic considerations incentivize \textcolor[rgb]{0.00,0.00,1.00}{Hank and/or Georgia} to move towards the center (from extremal positions of $h=1$ and $g=0$) to increase their influence over the public opinion. More broadly, we explore the following questions in this paper:
\begin{description}
  \item[Q1:] What are the properties of Nash equilibrium? Is it unique? Does it exist in pure or mixed strategies?
  \item[Q2:] How does CB impact equilibrium achieving strategies?
  \item[Q3:] Does it effect both \textcolor[rgb]{0.00,0.00,1.00}{Hank and Georgia} symmetrically in the sense that they move to center at equal amounts at the equilibrium? Do they move simultaneously, or only one of them moves?
  \item[Q4:] What are the effects of innate opinions on the Nash equilibrium in the presence of CB?
\end{description}

Before formulating the competitive game, let us first recall a technical lemma regarding the eigenvector and eigenvalue of network adjacency matrix.
\textcolor[rgb]{0.00,0.07,1.00}{\begin{lem}~\cite{spielman2009spectral}
Let $\mathfrak{G} = (\mathbb{V}, \mathbb{E})$ be a connected weighted graph. Assume there is a positive vector $\bar{{c}}$ such that the adjacency matrix $A$ satisfies $A \bar{ {c}} = \bar{\lambda} \bar{{c}}$. Then,  ${\bar{\lambda} = \mathop {\max }\limits_{i \in \mathbb{V}} \left\{ {\left|{\lambda _i}(A)\right|} \right\}}$ and the eigenvalue has  multiplicity 1.  \label{lem1}
\end{lem}}

\vspace{-0.40cm}
\textcolor[rgb]{0.00,0.00,1.00}{\begin{rem}
Lemma \ref{lem1} is a consequence of Perron-Frobenius theorem on non-negative matrices \cite{meyer2000matrix}. We note that Perron-Frobenius theorem requires the adjacency matrix $A$ to be irreducible, i.e., the implicit digraph $\mathfrak{G}$ must be strongly connected, while Lemma \ref{lem1} removes this strict requirement such that the digraph $\mathfrak{G}$ can be weakly connected. This is the motivation behind \eqref{adjpo} in Assumption \ref{assum1}.
\end{rem}}

We formulate the problem as a zero-sum game between \textcolor[rgb]{0.00,0.00,1.00}{Hank and Georgia}. As the cost function, by Lemma \ref{lem1} we use out-eigenvector centrality weighted cost, i.e.,
\begin{align}
f(g,h) = {c}^\top{x^*}(g,h), \label{eq:cob}
\end{align}
where ${x^*}(g,h)$ is computed via \eqref{eq:mmeq}, and $c = [c_{1},c_{2}, \ldots, c_{n}]^{\top} \in \mathbb{R}^{n}$ is the eigenvector associated with the largest eigenvalue of $W^{\top}$, i.e.,
\begin{align}
{W^ \top }c = \lambda c,\hspace{1.5cm}\textcolor[rgb]{0.00,0.00,1.00}{\lambda  = \mathop {\max }\limits_{i \in \mathbb{V}} \left\{ {\left|{\lambda _i}(W)\right|} \right\}}.
\label{eq:egceasso}
\end{align}

\vspace{-0.40cm}
\textcolor[rgb]{0.00,0.00,1.00}{\begin{rem}
The cost function $f(g,h) = {\mathbf{1}}^\top{x^*}(g,h)$ in~\cite{dhamal2018optimal} indicates that the decision maker treats individuals' opinions equally, which however does not hold in many  real social examples. For example, in the United States Electoral College, the number of each state's electors is equal to the sum of the state's membership in the Senate and House of Representatives, while in a company, the CEO usually has larger decision-making power than managers. Motivated by this observation, we assign relative scores to all individuals in a network based on the concept that the high-scoring individual contributes more influence to the decision making than the low-scoring individual. The relations in \eqref{eq:egceasso} indicate that $c$ in the cost function \eqref{eq:cob} is thus referred to the vector of out-eigenvector centralities that measures the importance of an individual in influencing other individuals' opinions \cite{newman2018networks}.\end{rem}
}

Here,  \textcolor[rgb]{0.00,0.07,1.00}{Hank's} objective is to maximize $f(g,h)$, while \textcolor[rgb]{0.00,0.00,1.00}{Georgia's} objective is to minimize $f(g,h)$. We next define two different notions:
\begin{align}
\widehat{s}  \triangleq \sum\limits_{i \in \mathbb{V}} {{{\widehat c}_i}} {s_i},\quad
\chi \triangleq \sum\limits_{i \in \mathbb{V}} {{{\widehat c}_i}\sum\limits_{j \in \mathbb{V}} {{s_i}{w_{ij}}} },\quad  \textcolor[rgb]{0.00,0.00,1.00}{{{\widehat c}_i} \triangleq \frac{{{c_i}}}{{\sum\limits_{j \in V} {{c_j}} }}}.
 \label{mean}
\end{align}
We note that $\widehat{s}$ represents the eigen-centrality weighted average of innate opinions over the network. In the special case of identical, neutral innate opinions, i.e.,  $s_i=1/2$ for all $i$, $\widehat{s}=1/2$ regardless of the network parameters. We also note that $\chi$ denotes a weighted average of short-term influence factored innate opinions, where weights are again eigen-centrality parameters. In the same aforementioned special case, i.e, $s_i=1/2$ for all $i$, it follows from \eqref{eq:egceasso} that $\chi  = \frac{{\sum\limits_{i \in \mathbb{V}} {\sum\limits_{j \in \mathbb{V}} {{c_i}} } {w_{ij}}}}{{2\sum\limits_{l \in \mathbb{V}} {{c_l}} }} = \frac{{{c^\top}W\mathbf{1}}}{{2{c^\top}\mathbf{1}}} = \frac{{\lambda {c^\top}\mathbf{1}}}{{2{c^\top}\mathbf{1}}} = \frac{\lambda }{2}$, which holds regardless of the remaining network parameters.

With these definitions,  we express the cost function as a function of $g$ and $h$ in the following corollary whose proof appears in Appendix C.

\begin{cor}
The cost function \eqref{eq:cob} can be expressed as:
\begin{align}
&f(g,h) \nonumber\\
& =\!\frac{((1 \!-\!2\beta  \!+\! (h \!-\! g)\gamma )\widehat{s} \!-\! \chi\!+\!(h\!+\! g)\beta\!+\!(g^2 \!-\! h^2)\gamma }{1 \!-\! \lambda  \!+\!(g\!-\!h)\gamma } c^ \top\! {\bf 1}. \label{eopo}
\end{align}
\label{lemxx2}
\end{cor}

This game can be viewed from two different perspectives, each of which provides a lower/upper bound for the value of the game $f(g^*, h^*)$.   The first one is a max-min optimization problem for \textcolor[rgb]{0.00,0.07,1.00}{Hank} $\mathop {\max }\limits_{h} \mathop {\min }\limits_{g} \left\{ f\left( {g,h} \right) \right\}$, where \textcolor[rgb]{0.00,0.07,1.00}{Hank} expresses her opinion $h$ to maximize $f(g,h)$, anticipating the rational best response of \textcolor[rgb]{0.00,0.00,1.00}{Georgia} ${g^*}\left( h \right)$, as formally stated below:
\begin{subequations}
  \begin{align}
{g^*}\left( h \right) &\triangleq \mathop {{\mathop{\rm argmin}\nolimits} }\limits_{g \in [0,\underline{s}]} \left\{ {f\left( {g,h} \right),~\text{for all}~h \in [{{\overline{s}},1} ]} \right\}, \label{eq:aie1}\\
{h^*} &\triangleq \mathop {{\mathop{\rm argmax}\nolimits} }\limits_{h \in [{{\overline{s}},1} ]} \left\{ {f\left( {{g^*}\left( h \right),h} \right)} \right\}.\label{eq:aie2}
\end{align}\label{eq:sdwb}
 \end{subequations}
\!\!\!\!\!Similarly,  a min-max optimization for \textcolor[rgb]{0.00,0.00,1.00}{Georgia} would be  $\mathop {\min }\limits_{g} \mathop {\max }\limits_{h} \left\{ f\left( {g,h} \right) \right\}$. In this scenario, \textcolor[rgb]{0.00,0.00,1.00}{Georgia} acts as the leader, with the objective to minimize $f(g, h)$ while taking the best response of \textcolor[rgb]{0.00,0.00,1.00}{Hank} into account. The strategies are referred to the pair $(h^*, g^{*})$ such that
\begin{subequations}
\begin{align}
{h^*}\left( g \right) &\triangleq \mathop {{\mathop{\rm argmax}\nolimits} }\limits_{h \in [\overline{s},1]} \left\{ {f({g,h}),~\text{for}~g \in [0, \underline{s} ]} \right\}, \label{eq:iae1}\\
{g^*} &\triangleq \mathop {{\mathop{\rm argmin}\nolimits} }\limits_{g \in [0,{{\underline{s}}} ]} \left\{ {f({g, {h^*}(g )})} \right\}.\label{eq:iae2}
\end{align}\label{eq:iae}
\end{subequations}
\vspace{-0.20cm}


The Nash equilibrium, if it exists in pure strategies, would be obviously the point, where $g^*, h^*$ that simultaneously satisfy \eqref{eq:aie2} and \eqref{eq:iae2}.

In next section, we formally show that this  game indeed admits a unique, pure-strategy Nash equilibrium, and hence, solving one of these optimization problems would be sufficient to derive the equilibrium-achieving strategies $g^*$ and $h^*$.



\section{Nash Equilibrium}
\textcolor[rgb]{0.00,0.00,1.00}{We first recall the definition of strategic form game, which will be used to investigate the properties of Nash Equilibrium.
\begin{defa}~\cite{exex}
A strategic form game is a triplet $\left\langle {\mathbb{I}, {\left( {{\mathbb{A}_i}} \right)}_{i \in \mathbb{I}} ,{\left( {{u_i}} \right)}_{i \in \mathbb{I}}} \right\rangle$, where
\begin{itemize}
  \item $\mathbb{I}$ is a finite set of players.
  \item $\mathbb{A}_i$ is a non-empty set of available actions for player $i$.
  \item $u_{i}: \mathbb{A} \rightarrow \mathbb{R}$ is the cost function of player $i$, where $\mathbb{A} = \prod\nolimits_{i \in \mathbb{I}} {{\mathbb{A}_i}}$.
\end{itemize}\label{defaaa}
\end{defa}}
\textcolor[rgb]{0.00,0.00,1.00}{We then transform the zero-sum games \eqref{eq:sdwb} and \eqref{eq:iae} to a strategic form game:} $\left\langle {\mathbb{I}, {\left( {{\mathbb{A}_i}} \right)}_{i \in \mathbb{I}} ,{\left( {{u_i}} \right)}_{i \in \mathbb{I}}} \right\rangle$, where
\begin{subequations}
\begin{align}
&\mathbb{I} \!=\! \{\text{\textcolor[rgb]{0.00,0.00,1.00}{Hank}}, \text{\textcolor[rgb]{0.00,0.00,1.00}{Georgia}}\}, ~\mathbb{A}_\text{\textcolor[rgb]{0.00,0.00,1.00}{Georgia}} \!=\! [0, \underline{s}], ~\mathbb{A}_\text{\textcolor[rgb]{0.00,0.00,1.00}{Hank}} \!=\! [\overline{s}, 1],\\ &u_{\text{\textcolor[rgb]{0.00,0.07,1.00}{Georgia}}}(a_{\text{\textcolor[rgb]{0.00,0.00,1.00}{Georgia}}}, a_{\text{\textcolor[rgb]{0.00,0.00,1.00}{Hank}}}) = -f(g,h), \\ &u_{\text{\textcolor[rgb]{0.00,0.00,1.00}{Hank}}}(a_{\text{\textcolor[rgb]{0.00,0.00,1.00}{Hank}}}, a_{\text{\textcolor[rgb]{0.00,0.00,1.00}{Georgia}}}) = f(g,h).
\end{align}\label{stragaaa}
\end{subequations}

\subsection{Existence and Uniqueness of Nash equilibrium}
\vspace{-0.00cm}
We start with the properties of  Nash equilibrium. Note that this equilibrium is unique and it exists only in pure strategies, i.e., there is no equilibrium in mixed strategies. This result is formally stated in the following theorem whose proof is presented in Appendix D.
\begin{thm}
The Nash equilibrium $(g^*, h^*)$ is unique and it is in pure strategies.
\label{exth}
\end{thm}

We proceed with the analysis of the aforementioned Nash equilibrium in the absence, and in the presence of CB.

\vspace{-0.40cm}
\subsection{Nash Equilibrium without CB: $\gamma = 0$}\label{nocbba}
We start with the case of no CB for which we have the intuitive solution, $(g^*,h^*)  = (0,1)$, regardless of the remaining problem parameters. We note that in our model setting $\gamma = 0$ in \eqref{eq:sdw} yields 'the no CB scenario'. Our result is stated formally in the following theorem whose proof is given in Appendix E.
\begin{cor}
When no individual holds CB towards the opinions of information sources \textcolor[rgb]{0.00,0.00,1.00}{Hank and Georgia}, the pure Nash equilibrium of the competitive information spread problems \eqref{eq:sdwb} and \eqref{eq:iae} is $(g^*,h^*)  = (0,1)$. \label{corfn}
\end{cor}

\vspace{-0.40cm}
\subsection{Nash Equilibrium with CB ($\gamma \ne 0$)}\label{nocb}
Before stating our main result, we define two simple functions ($q,m$) that are related to the cost function $f(g,h)$, and are used in the description of the equilibrium.
\begin{align}
q({g,h}) &\triangleq (\beta  + \gamma\widehat{s} - 2 \gamma h){a_1} + {b_1}\gamma  + {\gamma ^2}{h^2}, \label{eq:th1}\\
m( {g,h}) &\triangleq ( {\beta  - \gamma \widehat s + 2\gamma g}){a_2} + {g^2}{\gamma ^2} - {b_2}\gamma,\label{eq:defdef}
\end{align}
with
\begin{align}
{a_1} &= 1 - \lambda  + \gamma g, \label{eq:nd3}\\
{b_1} &= (1 - 2\beta - g\gamma)\widehat{s} - \chi + g\beta + g^{2}\gamma, \label{eq:nd5}\\
{a_2} &= 1 - \lambda  - h\gamma, \label{eq:nd4}\\
{b_2} &= \left( {1 - 2\beta  + h\gamma } \right)\widehat s + h\beta  - {h^2}\gamma  - \chi.\label{eq:nd6}
\end{align}
These functions are related to the partial derivates of the cost function as follows:
 \begin{align}
\frac{{\partial f\!\left( {g,h} \right)}}{{\partial h}} &= \frac{{q\!\left( {g,h} \right){c^\top }{{\bf{1}}}}}{{{{\left( {{a_1} - \gamma h} \right)}^2}}}, \label{eq:nda2}\\
\frac{{\partial f\!\left( {g,h} \right)}}{{\partial g}} &= \frac{m\!\left( {g,h} \right){c^\top{\bf{1}} }}{{{{\left( {{a_2} + g\gamma } \right)}^2}}}. \label{eq:nda1}
\end{align}

We next define the following auxiliary functions:
\begin{align}
&r(g) \!\triangleq\! \!- \frac{\sqrt {\!({1 \!-\! \lambda})\!({1 \!-\! \lambda  \!-\! \beta \!+\! 2g\gamma }) \!\!-\!\! ({2 \!-\!\lambda \!-\! 2\beta })\gamma\widehat{s} \!-\! 2g\beta\gamma \!+\! \gamma\chi}}{\gamma}\nonumber\\
&\hspace{1.0cm} + \frac{{1 - \lambda }}{\gamma } + g \label{eq:defeq2}\\
&w(h) \!\triangleq\! \frac{{\sqrt {\!(1 \!-\! \lambda)\!(1 \!-\! \lambda  \!-\! 2h\gamma  \!-\! \beta) \!\!+\!\! (2 \!-\! \lambda  \!-\! 2\beta)\gamma \widehat{s} \!+\! 2h\beta \gamma  \!-\! \chi\gamma } }}{\gamma } \nonumber\\
&\hspace{1.1cm}- \frac{{1 \!-\! \lambda}}{\gamma} + h. \label{dtm}
\end{align}
With these definitions at hand, we present the Nash equilibrium in the following theorem, and its proof in Appendix F.
\begin{thm}
The Nash equilibrium $(g^*, h^*)$ for \eqref{eq:sdwb} and \eqref{eq:iae} is:
\begin{itemize}
  \item If $m({0,1}) \ge 0$,
     \begin{align}
 (g^*,h^*)  = \begin{cases}
		(0,1), &\text{if } q(0,1) \ge 0\\
		(0,\overline{s}), &\text{if } q(0,\overline{s}) \le 0\\
		(0,r(0)), &\text{otherwise}.
	\end{cases}\label{ma1}
  \end{align}
  \item If $m(\underline{s},1) \le 0$,
     \begin{align}
      (g^*,h^*) = (\underline{s},1). \label{ma2}
  \end{align}
  \item Otherwise,
 \begin{align}
(g^*,h^*) = (w(1),1). \label{ma3}
  \end{align}
\end{itemize}
\label{thm:ai}
\end{thm}
\begin{rem}
By comparing \eqref{ma1} with \eqref{ma2} and \eqref{ma3}, we conclude that if $m({0,1}) \ge 0$, \textcolor[rgb]{0.00,0.00,1.00}{Georgia}'s pure strategy is fixed as $g^* = 0$; otherwise, \textcolor[rgb]{0.00,0.00,0.00}{Hank}'s strategy is fixed as $h^* = 1$.
\end{rem}

\begin{rem}
A rather interesting observation here is the following:  CB can influence only \textcolor[rgb]{0.00,0.00,1.00}{Hank's or Georgia's} pure strategy, while it cannot influence both of them simultaneously. In other words, either \textcolor[rgb]{0.00,0.00,1.00}{Hank or Georgia} moves to the center (or neither does so), but under no condition both \textcolor[rgb]{0.00,0.00,1.00}{Hank and Georgia} move towards the center at equilibrium.
\end{rem}

\vspace{-0.30cm}
\textcolor[rgb]{0.00,0.00,1.00}{\begin{rem}
Theorem \ref{thm:ai} indicates that the information sources need to explore the inference algorithms of social network topology, innate opinions and confirmation bias parameters for optimal information spread strategies. On the other hand, from perspective of security, Theorem \ref{thm:ai}
\begin{itemize}
  \item provides an optimal strategy to mitigate the influence of misinformation or disinformation on public opinions,
  \item implies that the defender can hinder the optimal information spread strategy of adversary information source through preserving privacy of partial network topology or some individuals' innate opinions from inference.
\end{itemize}
\end{rem}}

\vspace{-0.00cm}
\section{Impact Of CB and Innate Opinions}

\begin{figure*}
\centering
\subfigure{\includegraphics[height=2.5in,width=3.54in]{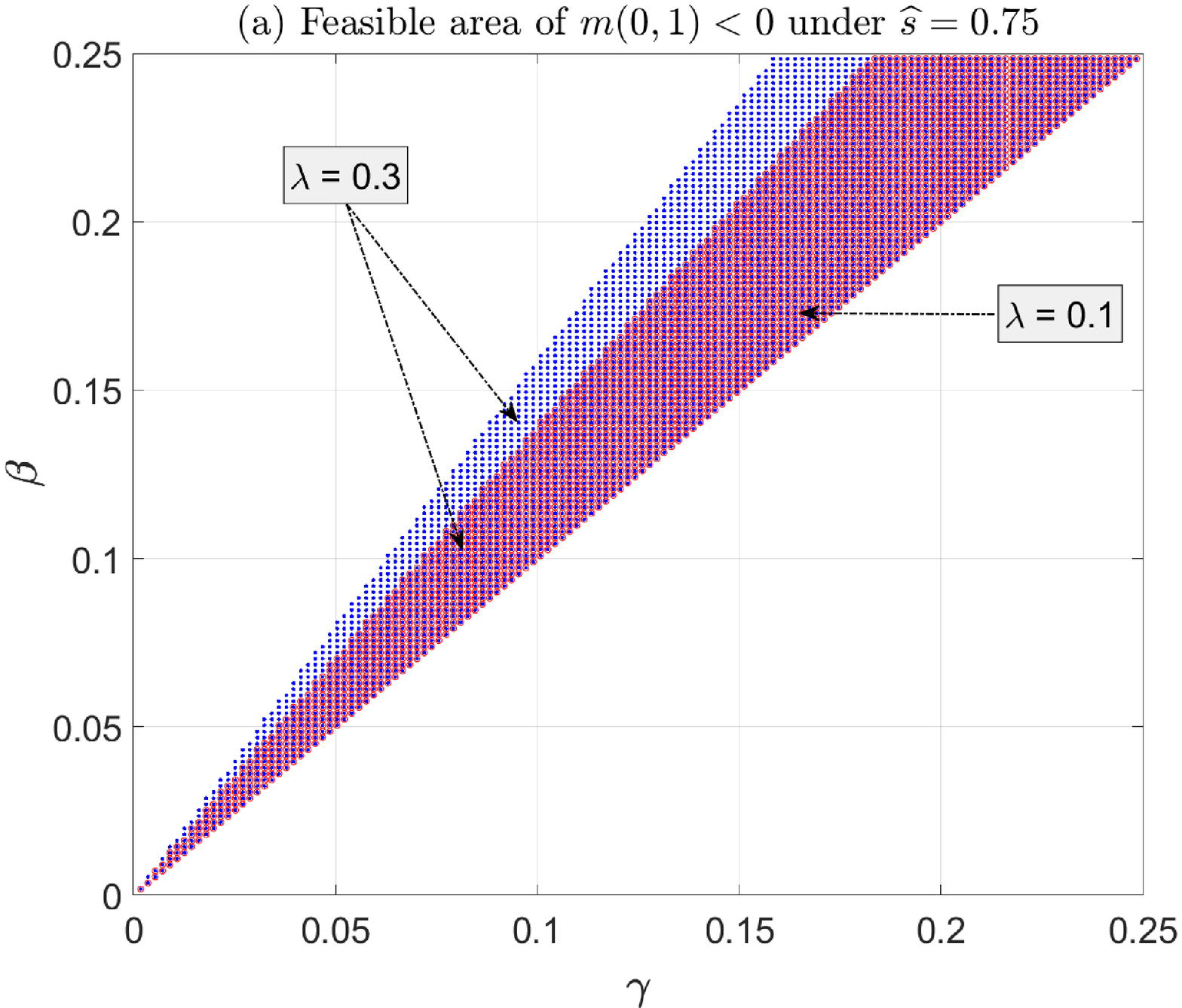}}
\subfigure{\includegraphics[height=2.5in,width=3.54in]{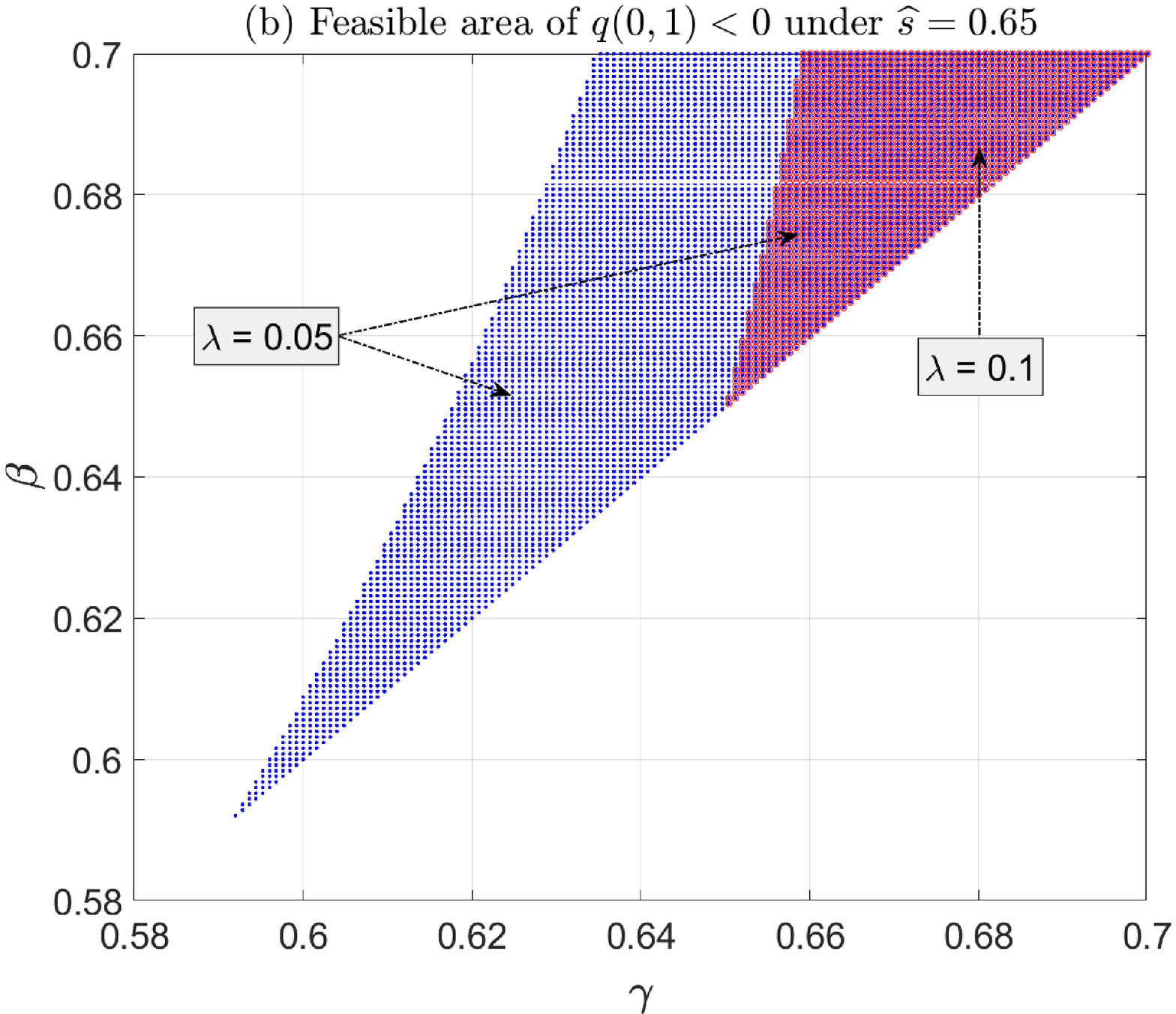}}
\caption{Under fixed $\xi = 0.14$: (a) feasible areas of $m(0,1) < 0$ under $(\widehat{s}, \lambda) = (0.75, 0.3)$ and $(\widehat{s}, \lambda) = (0.75, 0.1)$, (b) feasible areas of $q(0,1) < 0$ under $(\widehat{s}, \lambda) = (0.65, 0.05)$ and $(\widehat{s}, \lambda) = (0.65, 0.1)$.}
\label{fea}
\end{figure*}

Substituting $\gamma = 0$ into $q({0,1})$ and $m({0,1})$ straightforwardly yields $q({0,1}) \ge 0$ and $m({0,1}) \ge 0$, which, in conjunction with Theorem \ref{corfn} as well as the the conditions of \eqref{ma1}--\eqref{ma3}, indicate that the CB can influence the pure Nash equilibrium only when its parameters cause  $m({0,1}) < 0$, or $q({0,1}) < 0$. The feasible areas of \textcolor[rgb]{0.00,0.07,1.00}{$m({0,1}) < 0$ and $q({0,1}) < 0$} in each subfigure of Fig. \ref{fea} show that under fixed innate opinions and social network structure, there exists a range of $\beta$ and $\gamma$ in influencing the pure Nash equilibrium. The characterization of the conditions for which CB influences the Nash equilibrium is stated formally in the following theorem, whose proof is presented in Appendix G.
\begin{thm}
 CB changes the equilibrium-achieving strategy of \textcolor[rgb]{0.00,0.00,1.00}{Georgia}, i.e., $g^* \ne 0$, if and only if
\begin{align}
0 < \frac{1  -  \lambda   - 2\gamma  + 2\widehat{s}\gamma}{2\widehat{s} -\lambda \widehat{s} - \gamma - \chi} < \frac{\gamma}{\beta} \le 1, \label{bob}
\end{align}
and  CB alters \textcolor[rgb]{0.00,0.00,1.00}{Hank's} strategy, i.e., $h^* \ne 1$, if and only if
\begin{align}
0 <  \frac{1 - \lambda}{2 - 2\lambda  - 2\widehat{s} + \widehat{s}\lambda + 2\beta \widehat{s} + \chi  - \gamma }  < \textcolor[rgb]{0.00,0.00,1.00}{\frac{\gamma}{\beta}} \le 1. \label{Alice}
\end{align}
\label{influence}
\end{thm}

\textcolor[rgb]{0.00,0.00,1.00}{\begin{rem}
We verify from \eqref{eq:defdef} and \eqref{eq:th1} that \eqref{bob} and \eqref{Alice}, respectively, equate to $m(0,1) < 0$ and $q(0,1) < 0$. We  then obtain
\begin{align}
\frac{{\mathrm{d}m(0,1)}}{{\mathrm{d}\lambda }} = \widehat s\gamma  - \beta  \le 0, ~~\frac{{\mathrm{d}q(0,1)}}{{\mathrm{d}\lambda }} = 2\gamma  - \beta  - \gamma \widehat s, \nonumber
\end{align}
from which we observe that with $\beta > \gamma$, Georgia's equilibrium-achieving strategy is more likely to be changed by CB in a scenario of stronger social influence weights that result in larger $\lambda$ (which can be illustrated by the feasible areas of $m(0,1)$ in Fig.~\ref{fea} (a)). This observation does not hold for Hank since it depends on the weighted  average of innate opinions $\widehat{s}$, which can be demonstrated by Fig.~\ref{fea} that in contrast with feasible areas in Fig.~\ref{fea} (a), larger $\lambda$ does not lead to bigger feasible area of $q(0,1) < 0$.
\end{rem}}

\vspace{-0.40cm}
\textcolor[rgb]{0.00,0.00,1.00}{\begin{rem}
We note that when $\gamma = 0$, i.e., no individual holds confirmation bias, neither \eqref{bob} nor \eqref{Alice} holds, thus neither social influence nor innate opinion has influence on both Georgia's and Hank's equilibrium-achieving strategies.
\end{rem}}

\vspace{-0.40cm}
\subsection{Impact of Innate Opinion Distribution}
We next analyze the impact of the innate opinions on the equilibrium. The motivation to study the aforementioned impact stems from  comparing comparing Fig. \ref{fea} (a) and (b), which clearly demonstrate the importance of innate opinions on the critical regions for CB impact on the equilibrium-achieving strategies.


\subsubsection{Neutral Innate Opinions}
We first consider the setting where all innate opinions are neutral, i.e.,
\vspace{-0.10cm}
\begin{align}
s_{1} = s_{2} = \ldots = s_{n} =  \frac{1}{2}.  \label{neui}
\end{align}
\vspace{-0.50cm}

As we stated in Section III, this setting yields $\widehat{s}  =  \frac{1}{2}$ and $\chi = \frac{\lambda}{2}$. Substituting these values  in $m(0,1)$ and $q(0,1)$ yields $m(0,1) = q(0,1) = (\beta - \gamma)(1  -  \lambda   - \gamma) \ge 0$ via \eqref{eq:hhk}. These considerations result in the following corollary to Theorem~\ref{thm:ai}.
\begin{cor} Under \eqref{neui}, the Nash equilibrium for \eqref{eq:sdwb} and \eqref{eq:iae} is $(g^*,h^*)  = (0,1)$.
\label{nercc}
\end{cor}

\subsubsection{Extremal Innate Opinions}
We refer to the following set of innate opinions as extremal:
\vspace{-0.20cm}
\begin{align}
s_{1} = s_{2} = \ldots = s_{n} =  0.  \label{ex1a}\\
s_{1} = s_{2} = \ldots = s_{n} =  1.  \label{ex2a}
\end{align}
\vspace{-0.60cm}

With the consideration of $\widehat{s}$ defined in \eqref{mean}, we note that \eqref{ex1a} indicates that $\widehat{s} = \underline{s} = \overline{s}  = 0$. We substitute them into $q(0,0)$, $q(0,1)$ and $r(0)$, and re-denote them by $\tilde{q}(0,0)$, $\tilde{q}(0,1)$ and $\tilde{r}(0)$ in this scenario:
\begin{align}
\tilde{q}(0,0) &\!\triangleq\! \beta  \!-\! \beta \lambda  \!-\! \chi \gamma, ~~~~~\tilde{q}(0,1) \!\triangleq\! ( {\gamma  \!-\! \chi })\gamma  \!+\! (\beta  \!-\! 2\gamma )(1 \!-\! \lambda),\nonumber\\
\tilde{r}(0) &\!\triangleq\! \frac{{1 \!-\! \lambda  \!-\! \sqrt {(1 \!-\! \lambda )(1 \!-\! \lambda  \!-\! \beta) \!+\! \chi \gamma } }}{\gamma }.\nonumber
\end{align}
It can be verified from \eqref{eq:defdef} that under \eqref{ex1a} $m(0,1) \ge 0$ holds. Thus, we obtain the Nash equilibrium from Theorem~\ref{thm:ai}, which is formally stated in the following corollary.
\begin{cor}
Under \eqref{ex1a}, the Nash equilibrium for \eqref{eq:sdwb} and \eqref{eq:iae} is
\begin{align}
(g^*,h^*)  = \begin{cases}
		(0,1), &\text{if } \tilde{q}(0,1) \ge 0\\
		(0,0), &\text{if } \tilde{q}(0,0) \le 0\\
		(0,\tilde{r}(0)), &\text{otherwise}.
	\end{cases}\nonumber
\end{align}
\label{corcor1}
\end{cor}
\vspace{-0.20cm}

For \eqref{ex2a}, considering \eqref{mean} we have $\widehat{s} = \underline{s} = \overline{s}  = 1$. Substituting  these variables into $m(0,1)$, $m(1,1)$ and $w(0)$, we have \vspace{-0.20cm}
\begin{align}
\hat{m}(1,\!1) &\!\triangleq\! \chi \gamma  \!+\! \beta  \!-\! \beta \lambda  \!-\! \lambda \gamma\!,~\hat{m}(0,\!1) \!\triangleq\! \beta  \!-\! \lambda \beta  \!+\! (\lambda  \!+\! \gamma  \!+\! \chi  \!-\! 2)\gamma\!, \nonumber\\
\hat{w}(1) &\!\triangleq\! \frac{{\lambda  \!+\! \gamma  \!-\! 1 \!+\! \sqrt {1 \!-\! \beta  \!+\! (\lambda  \!+\! \gamma  \!+\! \beta  \!-\! 2)\lambda  \!-\! \chi \gamma } }}{\gamma }.\nonumber
\end{align} Moreover, in this scenario, we verify that $q(0,1) \ge 0$. By Theorem~\ref{thm:ai}, we obtain the following result.
\begin{cor}
Under \eqref{ex2a}, the Nash equilibrium for \eqref{eq:sdwb} and \eqref{eq:iae} is
\begin{align}
(g^*,h^*)  = \begin{cases}
		(0,1), &\!\!\text{if } \hat{m}(0,1) \!\ge\! 0\\
		(1,1), &\!\!\text{if } \hat{m}(1,1) \!\le\! 0\\
		(\hat{w}(1),1), &\!\!\text{otherwise}.
	\end{cases}\nonumber
\end{align}
\label{corcor33}
\end{cor}

\vspace{-0.35cm}
\section{Numerical Results}
\vspace{-0.10cm}
\begin{figure}[http]
\centering
\includegraphics[scale=0.42]{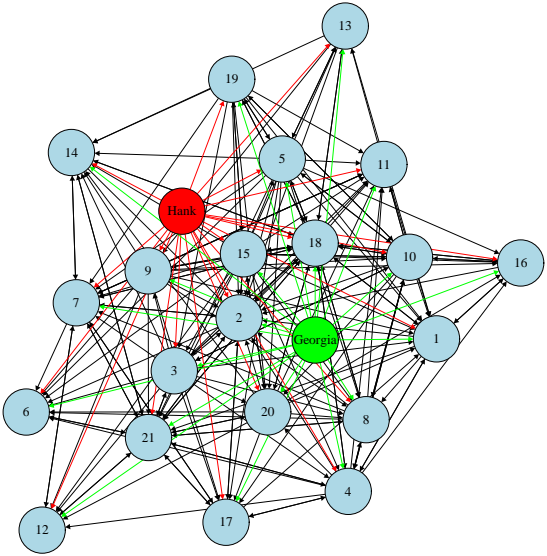}
\caption{Krackhardt's advice network~\cite{krackhardt1987cognitive} in the presence of competitive information sources \textcolor[rgb]{0.00,0.00,1.00}{Hank and Georgia}.}
\label{fig:lmdd}
\end{figure}

In this section, we numerically demonstrate our results in the well-known Krackhardt's advice network~\cite{krackhardt1987cognitive} with 21 individuals. The network topology is shown in Fig.~\ref{fig:lmdd}, where \textcolor[rgb]{0.00,0.00,1.00}{Hank and Georgia} represent two competitive information sources. For the weight matrix $W$, if individual $\mathrm{v}_{i}$ asks for advice from her neighbor $\mathrm{v}_{j}$, then $w_{ij} = \frac{1}{25 + \Gamma_{i}^{\emph{\emph{in}}}}$ for all the individuals $\mathrm{v}_{j}$ that influence individual $\mathrm{v}_{i}$,  where ${{\Gamma}^{\emph{\emph{in}}}_i}$ denotes in-degree of individual $\mathrm{v}_{i}$. The largest eigenvalue of the adjacency matrix $W$ that describes the structure of Krackhardt's advice network is computed as $\lambda =  0.2369$.

In this section, under the fixed social network structure, we demonstrate the five different Nash equilibria presented in Theorem \ref{thm:ai} through setting different groups of innate opinions and CB parameters.

\begin{figure*}[!t]
\centering
\subfigure{\includegraphics[height=2.00in,width=2.35in]{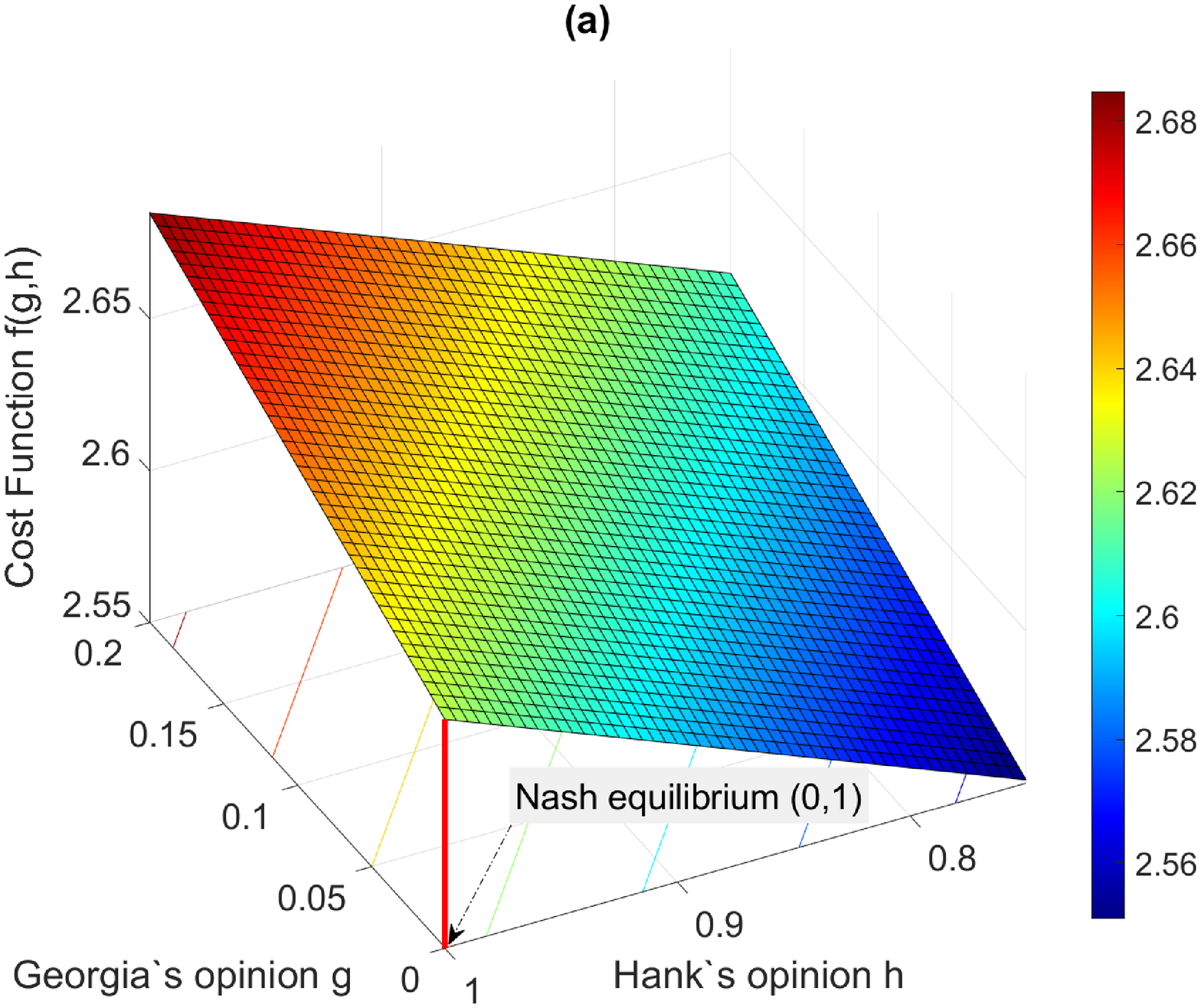}}
\subfigure{\includegraphics[height=2.00in,width=2.35in]{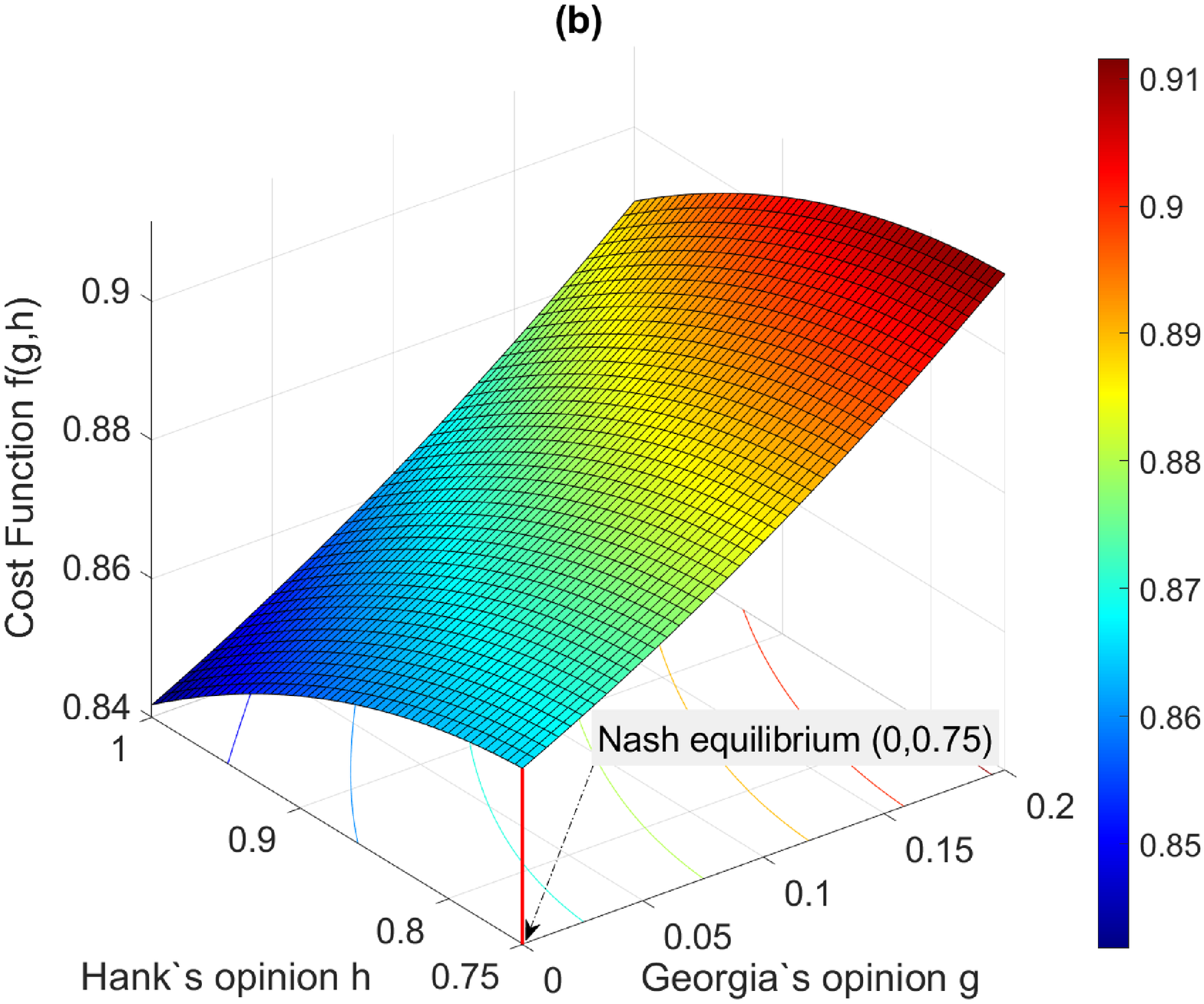}}
\subfigure{\includegraphics[height=2.00in,width=2.35in]{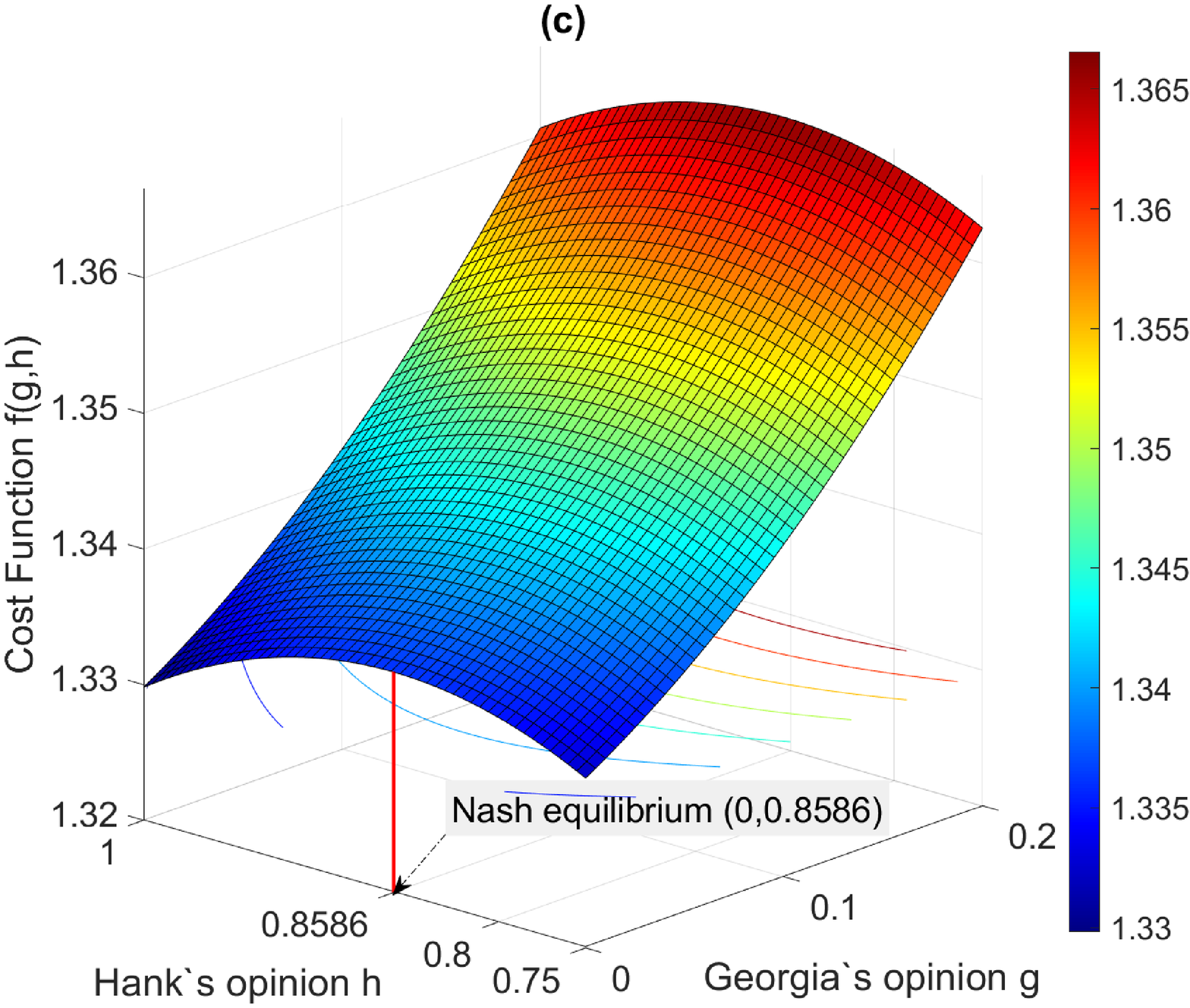}}
\subfigure{\includegraphics[height=2.00in,width=2.35in]{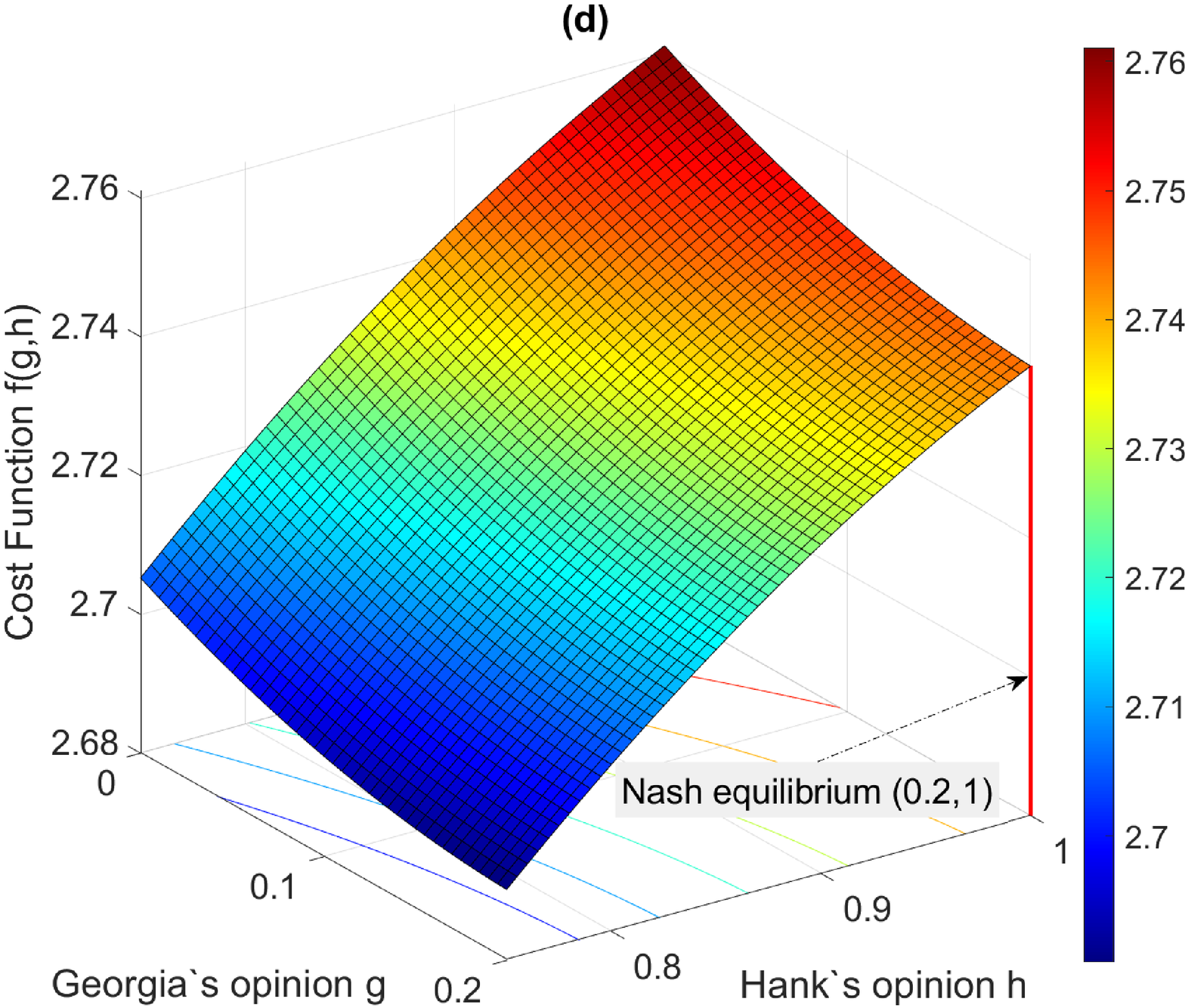}}
\subfigure{\includegraphics[height=2.00in,width=2.35in]{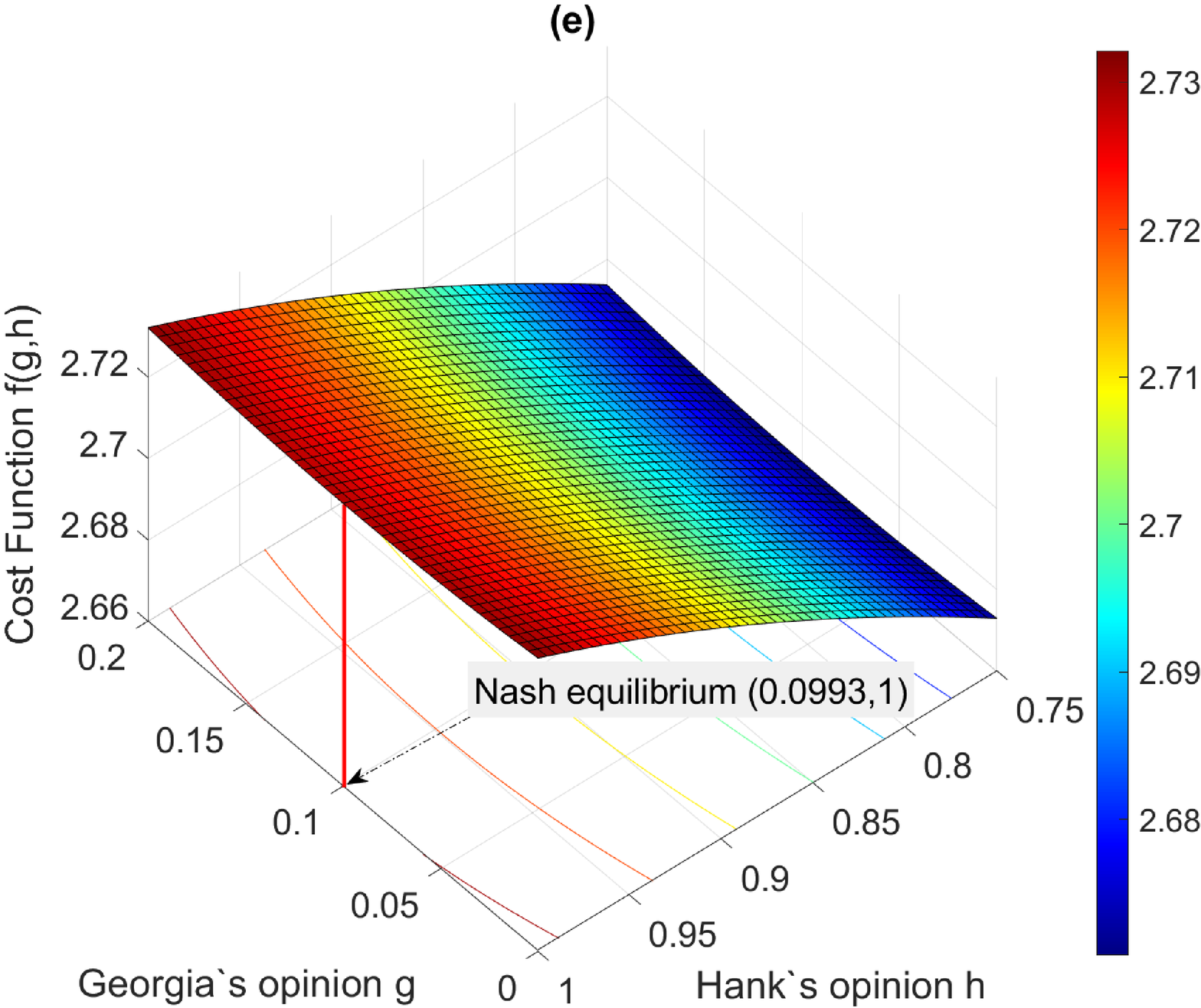}}
\caption{Surface plots of cost functions under different conditions: (a) no CB ($m(0,1) \geq 0$ and $q(0,1) \geq 0$), (b) $m(0,1) \geq 0$ and $q(0,\overline{s}) \leq 0$, (c)
$m({0,1}) \ge 0$ and $q(0,1) < 0$ and $q(0,\overline{s}) > 0$, (d) $m(\underline{s},1) \le 0$, (e) $m({0,1}) < 0$ and $m(\underline{s},1) > 0$.}
\label{sp}
\end{figure*}

\vspace{-0.35cm}
\subsection{No CB}
\vspace{-0.05cm}
We let the innate opinions of individuals $\mathrm{v}_{1}$ and $\mathrm{v}_{2}$ be  the same as 0.2, while others are uniformly set as 0.75. For CB, we set $\beta = 0.06$ and $\gamma = 0$, which indicates that no individual holds CB toward the opinions of \textcolor[rgb]{0.00,0.00,1.00}{Hank and Georgia}. We verify that $q(0,1) \geq 0$ and $m(0,1) \geq 0$. By Corollary \ref{corfn} or Theorem \ref{thm:ai}, we theoretically expect the Nash equilibrium to be $(g^{*},h^{*}) = (0,1)$, which is numerically demonstrated by Fig.~\ref{sp} (a).

\vspace{-0.38cm}
\subsection{$m(0,1) \geq 0$ and $q(0,\overline{s}) \leq 0$}
\vspace{-0.10cm}
We set individual $\mathrm{v}_{21}$'s innate opinion  as 0.2, while others are uniformly set as 0.75.  For CB, we let $\beta = \gamma = 0.06$. Under this setting, we have
\begin{align}
\underline{s} = 0.2, ~\overline{s} = 0.75, ~\widehat{s} =  0.2283, ~\chi = 0.0580, \nonumber
\end{align}
by which, we  verify from \eqref{eq:th1}--\eqref{eq:nd6} that $m(0,1) \geq 0$ and $q(0,\overline{s}) \leq 0$. Hence, from Theorem \ref{thm:ai} we expect the Nash equilibrium to be $(g^{*},h^{*}) = (0,\overline{s}) = (0,0.75)$, which is demonstrated by Fig.~\ref{sp} (b).

\vspace{-0.50cm}
\subsection{$m({0,1}) \ge 0$ and $q(0,1) < 0$ and $q(0,\overline{s}) > 0$}
\vspace{-0.10cm}

We let the innate opinions of individuals $\mathrm{v}_{18}-\mathrm{v}_{21}$ be  the same as 0.75,  \textcolor[rgb]{0.00,0.07,1.00}{while others are uniformly set as 0.2}. For CB, we choose $\beta = \gamma = 0.06$.
Under this setting, we have
\begin{align}
\underline{s} = 0.2, ~\overline{s} = 0.75, ~\widehat{s} =  0.3637, ~\chi = 0.0945, \nonumber
\end{align}
by which, we  verify from \eqref{eq:th1}--\eqref{eq:nd6} that $m({0,1}) \ge 0$ and $q(0,1) < 0$ and $q(0,\overline{s}) > 0$. Moreover, we obtain from \eqref{eq:defeq2} that $r(0) = 0.8586$. Therefore, from Theorem \ref{thm:ai} we expect the Nash equilibrium to be $(g^{*},h^{*}) = (0,r(0)) = (0,0.8586)$, which  is demonstrated by Fig.~\ref{sp} (c).

\vspace{-0.45cm}
\subsection{$m(\underline{s},1) \le 0$}
\vspace{-0.08cm}
We set the innate opinions of individuals $\mathrm{v}_{1}$ and $\mathrm{v}_{2}$ as the same as 0.2, others are uniformly set as 0.75. For CB, we choose $\beta = \gamma = 0.06$.
Under this setting, we have
\begin{align}
\underline{s} = 0.2, ~\overline{s} = 0.75, ~\widehat{s} =  0.7265, ~\chi = 0.1693, \nonumber
\end{align}
by which, we  verify from \eqref{eq:th1}--\eqref{eq:nd6} that $m(\underline{s},1) \le 0$. From Theorem \ref{thm:ai} we expect Nash equilibrium to be $(g^{*},h^{*}) = (\underline{s}, 1) = (0.2,1)$, which  is demonstrated by Fig.~\ref{sp} (d).

\vspace{-0.40cm}
\subsection{$m({0,1}) < 0$ and $m(\underline{s},1) > 0$}
In this case, we choose the same setting of innate opinions in case D, but for CB, we let $\beta = 0.06$ and $\gamma = 0.048$.  We verify from \eqref{eq:th1}--\eqref{eq:nd6} that $m({0,1}) < 0$ and $m(\underline{s},1) > 0$. Moreover, by \eqref{dtm} we have $w(1) = 0.0993$. Hence, from Theorem \ref{thm:ai} we expect the Nash equilibrium to be $(g^{*},h^{*}) = (w(1), 1) = (0.0993,1)$, which  is demonstrated by Fig.~\ref{sp} (e).

\vspace{-0.20cm}
\section{Conclusion}
\vspace{-0.10cm}

In this paper, we have studied the competitive information spread with confirmation bias (CB) over social networks, which is formulated as a zero-sum game and have investigated the pure Nash equilibrium point. We have analyzed the impact of CB and innate opinions of the Nash equilibrium, particularly the following trade-off for information sources: a move to the extremal opinions to maximally change public opinion, and another move, due to the existence CB, to the center to maximize the influence. Our analysis has uncovered a few rather surprising results: CB moves the Nash equilibrium towards the center only when the innate opinions are not neutral, and this move occurs for only one of the information sources. Theoretical results are verified by numerical examples.

There are several directions for future work; some of which are  listed as follows:
\begin{itemize}
  \item Investigation of the inferences of group innate opinions, social network topology and parameters of confirmation bias.
  \item Incorporation of group centralities into the cost function to identify the critical group of individuals in information spreading.
\end{itemize}

\vspace{-0.15cm}
\appendices
\section*{Appendix A: Auxiliary Results}
\vspace{-0.05cm}
This section presents the auxiliary results for the proofs of main results.

\begin{lem}
The matrix $E$ defined in~(\ref{eq:def1}) satisfies:
\begin{align}
\frac{{c^\top E}}{{1 - \lambda  + \left( {g - h} \right)\gamma }} = c^\top.
\label{eq:egce}
\end{align}
\end{lem}

\begin{IEEEproof}
We note that $W$ is an adjacency matrix and its transposition does not change its eigenvalues; by Lemma~\ref{lem1} we have \eqref{eq:egceasso}. It follows from \eqref{eq:def1} and \eqref{eq:egceasso} that
\begin{align}
{{c}^ \top }E = {{c}^ \top } - {{c}^ \top }W + ({g - h}){{c}^\top }\gamma = ({1 - \lambda  +({g - h})\gamma}){{c}^ \top }, \nonumber
\end{align}
from which \eqref{eq:egce} is obtained.
\end{IEEEproof}

\begin{lem}
With $\gamma \ne 0$,  $q(g,h)$ and $m(g,h)$ in \eqref{eq:th1} and \eqref{eq:defdef} satisfy $m(g,h) + q(g,h) > 0$.\label{ladd1}
\end{lem}

\begin{IEEEproof}
The partial derivative of \eqref{eq:defdef} w.r.t. $g$ is
\begin{align}
\frac{{\partial m\!\left( {g,h} \right)}}{{\partial g}} = 2\gamma {a_2} + 2{\gamma ^2}g.\label{eq:defdefdef}
\end{align}
Noticing the eigenvalue $\lambda$ given in \eqref{eq:egceasso}, the condition \eqref{eq:hhk}, in conjunction with Ger$\check{\emph{\emph{s}}}$gorin disk theorem~\cite{horn1985c}, imply that
\begin{align}
1 - \lambda  \ge 1 - \max \left\{ {{{\left\| W \right\|}_1},{{\left\| W \right\|}_\infty }} \right\} > \max \left\{ {2\beta ,4\gamma } \right\},\label{eq:a2con}
\end{align}
which together with~(\ref{eq:nd4}) and the fact $0< h < 1$ imply that $a_{2} > 0$. Thus, we conclude from \eqref{eq:defdefdef} that
\begin{align}
\frac{{\partial m(g,h)}}{{\partial g}} & \ge 0, ~~\text{for} ~g, h \in [0,1]. \label{desg}
\end{align}
The partial derivative of \eqref{eq:defdef} w.r.t. $h$ satisfies:
\begin{align}
\frac{{\partial m(g,h)}}{{\partial h}} =  - 2(\beta  - \gamma (h - g) ) \gamma  \le 0, \text{for} ~g, h \in [0,1]. \label{desh}
\end{align}
Applying the same analysis to \eqref{eq:th1}, we have
\begin{align}
\frac{{\partial q( {g,h})}}{{\partial g}} & = 2(\beta \!+\! (g \!-\! h)\gamma)\gamma \!\ge\! 0, \text{for} ~g, h \!\in\! [0,1], \label{dvh1}\\
\frac{{\partial q( {g,h})}}{{\partial h}} & = -2(1 \!-\! \lambda \!+\! (g \!-\! h)\gamma)\gamma \!\le\! 0, \text{for} ~g, h \!\in\! [0,1]. \label{dvh2}
\end{align}
From \eqref{desg}-\eqref{desh} and  \eqref{dvh1}-\eqref{dvh2}, we have:
\begin{align}
m(g,h) \ge  m(0,h) \ge m(0,1), \label{rep1}\\
q(g,h) \ge  q(0,h) \ge  q(0,1). \label{rep2}
\end{align}
Combining \eqref{rep1} and \eqref{rep2} yields
\begin{align}
m(g,h) + q(g,h) \ge m(0,1) + q(0,1).\label{pladd11}
\end{align}
Substituting the values of $m(0,1)$ and $q(0,1)$ into \eqref{pladd11}, we have
\begin{align}
m(g,h) + q(g,h)  &\ge 2(1 - \frac{\lambda }{2} - \gamma  - \gamma \lambda)\beta  + 2{\gamma ^2} + 2\gamma \lambda \nonumber\\
&\ge 2(1 - \frac{\lambda }{2} - \gamma  - \gamma \lambda)\gamma  + 2{\gamma ^2} + 2\gamma \lambda \nonumber\\
&= 2\gamma  + \lambda \gamma  - 2{\gamma ^2}\lambda = 2\gamma (1 - \gamma \lambda ) + \lambda \gamma  \nonumber\\
&> 0, \nonumber
\end{align}
where the inequalities follow from $\gamma \ne 0$, \eqref{eq:asfx00} and \eqref{eq:a2con}.
\end{IEEEproof}

\begin{lem}
Consider $f({g,h})$, $q({g,h})$ and $m({g,h})$ given by \eqref{eopo}, \eqref{eq:th1} and  \eqref{eq:defdef}, respectively. If $q(g,h) \ge 0$, $m(\tilde{g},\tilde{h}) \ge 0$, $1 \ge g \ge \tilde{g} \ge 0$ and $1 \ge h \ge \tilde{h} \ge 0$, then $f(g,h) \ge f(\tilde{g},\tilde{h})$.\label{kkm1}
\end{lem}

\begin{IEEEproof}
It follows from \eqref{dvh2} that $q(g,h) \ge 0$ implies $q(g,\breve{h}) \ge 0$ for $\breve{h} \in [\tilde{h},h] \subseteq [0,1]$, which, in conjunction with  \eqref{eq:nda2}, results in
\begin{align}
f(g,h) \ge f(g,\tilde{h}). \label{xcv1}
\end{align}
Meanwhile, following \eqref{dvh1}, $m(\tilde{g},\tilde{h}) \ge 0$ implies $m(\breve{g},\tilde{h}) \ge 0$ for  $\breve{g} \in [\tilde{g}, g] \subseteq [0, 1]$,
which, in conjunction with  \eqref{eq:nda1}, results in $f(g,\tilde{h}) \ge f(\tilde{g},\tilde{h})$,  which along with \eqref{xcv1} leads to $f(g,h) \ge f(g,\tilde{h}) \ge f(\tilde{g},\tilde{h})$.
\end{IEEEproof}

\section*{Appendix B: Proof of Theorem~\ref{thm:th1}}
Noting that $s_{i}, x_{i}(0) \in [0,1]$, $\forall i \in \mathbb{V}$, in conjunction with \eqref{eq:as122f}, we obtain $x_{i}(k) \in [0,1]$. Thus, the non-negativeness of state-dependent influence weights \eqref{eq:sdw} directly follows from \eqref{eq:asfx00}. We denote the mapping executed by the dynamics in (1) from time $k$ to $k+1$ as $\Psi$, i.e., ${x_i}({k + 1})=\Psi_i(x_i(k))$. For two  vectors $x$ and $y$, we have
\begin{align}
&{\Psi_i}({{x_i}}) - {\Psi_i}({y_i}) \nonumber\\
&= ({{\alpha _i}({x_i}) \!-\! {\alpha _i}(y_i)}){s_i} \!+\! \sum\limits_{j \in \mathbb{V}} {{w_{ij}}({{x_j} \!-\! y_j})}  \nonumber\\
&\hspace{0.50cm}  + ({{\overline{w}}({x_i}) \!-\! {\overline{w}}(y_i)})h \!+\! ({{\underline{w}}({x_i}) \!-\! {\underline{w}}(y_i)})g, ~~i \in \mathbb{V}.  \label{nn4}
\end{align}
Also noting that
\begin{align}
\left| {{\alpha _i}({x_i}) \!-\! {\alpha _i}({y_i})} \right| &= \gamma \left( {\left| {{x_i} \!-\! h} \right| - \left| {{y_i} \!-\! h} \right| + \left| {{x_i} \!-\! g} \right| - \left| {{y_i} \!-\! g} \right|} \right) \nonumber\\
& \le 2\gamma \left| {{x_i} - {y_i}} \right|. \label{nn5}
\end{align}
Moreover, from \eqref{eq:sdw} we have
\begin{subequations}
\begin{align}
\left| {{{\overline{w}}}({x_i}) \!-\! {{\overline{w}}}({y_i})} \right| \!=\! \gamma \left| {\left| {{y_i} \!-\! h} \right| - \left| {{x_i} \!-\! h} \right|} \right| \!\le\! \gamma \left| {{x_i} \!-\! {y_i}} \right|, \\
\left| {{{\underline{w}}}({x_i}) \!-\! {{\underline{w}}}({y_i})} \right| \!=\! \gamma \left| {\left| {{y_i} \!-\! g} \right| - \left| {{x_i} \!-\! g} \right|} \right| \!\le\! \gamma \left| {{x_i} \!-\! {y_i}} \right|.
  \end{align}\label{nn6}
\end{subequations}
Combining \eqref{nn4}  with \eqref{nn5} and \eqref{nn6} yields
\begin{align}
\left\| {\Psi(x) - \Psi(y)} \right\| &= \sum\limits_{i \in \mathbb{V}} {\left| {{\Psi_i}({x_i}) - {\Psi_i}({y_i})} \right|} \nonumber\\
& \le 4\gamma \sum\limits_{i \in \mathbb{V}} {\left| {{x_i} - {y_i}} \right|}  + \sum\limits_{i \in \mathbb{V}} {\sum\limits_{j \in \mathbb{V}} {{w_{ij}}\left| {{x_j} - {y_j}} \right|} } \nonumber\\
& = 4\gamma \sum\limits_{i \in \mathbb{V}} {\left| {{x_i} - {y_i}} \right|}  + \sum\limits_{i \in \mathbb{V}} {\left| {{x_i} - {y_i}} \right|\sum\limits_{j \in \mathbb{V}} {{w_{ji}}} } \nonumber\\
& \le 4\gamma \sum\limits_{i \in \mathbb{V}}\!{\left| {{x_i} \!-\! {y_i}} \right|}  \!+\! \sum\limits_{i \in \mathbb{V}} \!{\left| {{x_i} \!-\! {y_i}} \right|\!\mathop {\max }\limits_{i \in \mathbb{V}}\{{\sum\limits_{j \in \mathbb{V}} {{w_{ji}}} }\}} \nonumber\\
& = \left( {4\gamma  + {{\left\| W \right\|}_1}} \right)\left\| {x - y} \right\|.
  \end{align}
Here we note that  if $1 - {\left\| W \right\|_{1} } > 4\gamma$, due to Banach fixed-point theorem (following the steps in the proof of Theorem 1 in~\cite{mao2018spread}), the dynamics in \eqref{eq:od} converge to a unique point, regardless of the initial state. This condition, in conjunction with \eqref{com1} yields \eqref{eq:asfx}.

To solve for $x^{*}(g,h)$,  we set $g\mathbf{1} \ge {x}(0) \ge h\mathbf{1}$ (since $x^{*}(g,h)$ is independent of the
initial state, we can set  arbitrary initial condition).  Using \eqref{eq:oofn} and \eqref{eq:as122f}, we have $h\mathbf{1} \ge {x}(k) \ge g\mathbf{1}$ for $\forall k \in \mathbb{N}_{0}$, and hence  $h\mathbf{1} \ge {x^{*}(g,h)} \ge g\mathbf{1}$. This implies, by re-expressing (\ref{eq:sdw}) for $x^*(g,h)$, that
   \begin{subequations}
  \begin{align}
\overline{w}(x^*_{i}(g,h)) = \beta  - h\gamma  + \gamma x_i^*(g,h), \label{eq: sdw1b}\\
\underline{w}(x^*_{i}(g,h)) = \beta  + g\gamma - \gamma x_i^*(g,h),\label{eq: sdw2b}
\end{align}\label{eq:sdwk}
 \end{subequations}
and also from~(\ref{eq:as122f}):
\begin{align}
{\alpha _i}(x^*_{i}(g,h)) = 1 - \sum\limits_{j \in \mathbb{V}}  {w_{ij}} - 2\beta  + \left( {h - g} \right)\gamma, i \in \mathbb{V}.\label{eq:resp}
\end{align}
Plugging (\ref{eq:sdwk}) and~(\ref{eq:resp}) in (\ref{eq:odd}), we obtain  (\ref{eq:mmeq}).

\section*{Appendix C: Proof of Corollary \ref{lemxx2}}
Plugging \eqref{eq:mmeq} and \eqref{eq:egce} into \eqref{eq:cob} yields, after some algebra:
\begin{align}
f(g,h) &\!=\! \frac{{\sum\limits_{i \in\mathbb{V}} {{c_i}} }}{{1 \!-\! \lambda  \!+\! (g \!-\! h)\gamma }}\left( {\sum\limits_{i \in \mathbb{V}}  ((h \!+\! g)\beta  \!+\! ({g^2} \!-\! {h^2})\gamma )\frac{{{c_i}}}{{\sum\limits_{i \in \mathbb{V}} {{c_i}} }}} \right.\nonumber\\
&\hspace{1.2cm}\left. { \!+\! \sum\limits_{i \in \mathbb{V}} {(1 \!-\! \sum\limits_{j \in \mathbb{V}} {{w_{ij}}}  \!-\! 2\beta  \!+\! (h \!-\! g)\gamma )\frac{{{c_i}{s_i}}}{{\sum\limits_{i \in \mathbb{V}} {{c_i}} }}} } \right),\nonumber
\end{align}
which is equivalent to \eqref{eopo},  considering \eqref{mean}.

\section*{Appendix D: Proof of Theorem \ref{exth}}
Let us consider the transformed strategic form game $\left\langle {\mathbb{I}, {\left( {{\mathbb{A}_i}} \right)}_{i \in \mathbb{I}} ,{\left( {{u_i}} \right)}_{i \in \mathbb{I}}} \right\rangle$ with elements given in \eqref{stragaaa}. We obtain from \eqref{eq:nda1} with \eqref{eq:defdef}, \eqref{eq:nd4}, \eqref{eq:nd6} and \eqref{stragaaa} that
\begin{align}
\frac{\partial^{2} u_{\text{Georgia}}(a_{\text{\textcolor[rgb]{0.00,0.00,1.00}{Georgia}}}, a_{\text{\textcolor[rgb]{0.00,0.00,1.00}{Hank}}})}{\partial a^{2}_{\text{\textcolor[rgb]{0.00,0.00,1.00}{Georgia}}}} \!=\! -\frac{{{\partial ^2}f( {g,h} )}}{{\partial {g^2}}} \!=\! - \frac{{2\gamma {c^ \top }{\bf{1}}\bar{\mathrm{m}}( {g,h} )}}{{{{( {{a_2} + g\gamma } )}^3}}}, \label{ppokx1}
\end{align}
where
\begin{align}
\bar{\mathrm{m}}\left( {g,h} \right) = &\left( {1 - \lambda  - h\gamma } \right)\left( {1 - \lambda  - h\gamma  - \beta  + \gamma \widehat{s}} \right) \nonumber\\
 &+ \left( {\left( {1 - 2\beta  + h\gamma } \right)\widehat{s} + h\beta  - {h^2}\gamma  - \chi } \right)\gamma. \label{ppokx2}
\end{align}
Noticing \eqref{mean}, we have $\chi  \le \widehat{s}$ (since $0 \le \sum\limits_{j \in \mathbb{V}} {{w_{ij}}}  \le 1$), moreover, considering \eqref{eq:oofn} we have $h \ge \widehat{s}$. We then obtain from \eqref{ppokx2} that
\begin{align}
\bar{\mathrm{m}}\left( {g,h} \right) \ge &\left( {1 - \lambda  - h\gamma } \right)\left( {1 - \lambda  - h\gamma  - \beta  + \gamma \widehat{s}} \right) \nonumber\\
& + \left( {\left( {1 - 2\beta  + \widehat{s}\gamma } \right)\widehat{s} + \widehat{s}\beta  - {\widehat{s}^2}\gamma  - \widehat{s}} \right)\gamma \nonumber\\
= &\left( {1 - \lambda  - h\gamma } \right)\left( {1 - \lambda  - h\gamma  - \beta  + \gamma  \widehat{s}} \right) - \beta \gamma \widehat{s},\nonumber
\end{align}
which, in conjunction with \eqref{eq:hhk}, further implies
\begin{align}
\bar{\mathrm{m}}\left( {g,h} \right) &\ge \left( {1 \!-\! \lambda  \!-\! h\gamma } \right)\left( {1 \!-\! \lambda  \!-\! h\gamma  \!-\! \beta  \!+\! \gamma \widehat{s}} \right) \!-\! \left( {1 \!-\! \lambda  \!-\! h\gamma } \right)\gamma \widehat{s}  \nonumber\\
 &= \left( {1 - \lambda  - h\gamma } \right)\left( {1 - \lambda  - h\gamma  - \beta } \right) \ge 0.\label{ppokx3}
\end{align}

Substituting \eqref{ppokx3} into \eqref{ppokx1} yields  $\frac{\partial^{2} u_{\text{\textcolor[rgb]{0.00,0.00,1.00}{Georgia}}}(a_{\text{\textcolor[rgb]{0.00,0.00,1.00}{Georgia}}}, a_{\text{\textcolor[rgb]{0.00,0.00,1.00}{Hank}}})}{\partial a^{2}_{\text{\textcolor[rgb]{0.00,0.00,1.00}{Georgia}}}} \le 0$. Following the same method, we also obtain $\frac{\partial^{2} u_{\text{\textcolor[rgb]{0.00,0.00,1.00}{Hank}}}(a_{\text{\textcolor[rgb]{0.00,0.00,1.00}{Georgia}}}, a_{\text{\textcolor[rgb]{0.00,0.00,1.00}{Hank}}})}{\partial a^{2}_{\text{\textcolor[rgb]{0.00,0.00,1.00}{Hank}}}} \le 0$. Thus, $\left\langle {\mathbb{I}, {\left( {{\mathbb{A}_i}} \right)}_{i \in \mathbb{I}} ,{\left( {{u_i}} \right)}_{i \in \mathbb{I}}} \right\rangle$ is a concave game.

Following the proof of Lemma \ref{kkm1} in Appendix A, we obtain $u_{\text{\textcolor[rgb]{0.00,0.00,1.00}{Georgia}}}(a_{\text{\textcolor[rgb]{0.00,0.00,1.00}{Georgia}}}, a_{\text{\textcolor[rgb]{0.00,0.00,1.00}{Hank}}}) \ge u_{\text{\textcolor[rgb]{0.00,0.00,1.00}{Georgia}}}(\bar{a}_{\text{\textcolor[rgb]{0.00,0.00,1.00}{Georgia}}}, a_{\text{\textcolor[rgb]{0.00,0.00,1.00}{Hank}}})$ and $u_{\text{\textcolor[rgb]{0.00,0.00,1.00}{Hank}}}(a_{\text{\textcolor[rgb]{0.00,0.00,1.00}{Georgia}}}, a_{\text{\textcolor[rgb]{0.00,0.00,1.00}{Hank}}}) \ge u_{\text{\textcolor[rgb]{0.00,0.00,1.00}{Hank}}}(\bar{a}_{\text{\textcolor[rgb]{0.00,0.00,1.00}{Georgia}}}, \bar{a}_{\text{\textcolor[rgb]{0.00,0.00,1.00}{Hank}}})$, for any $a_{\text{\textcolor[rgb]{0.00,0.00,1.00}{Georgia}}} < \bar{a}_{\text{\textcolor[rgb]{0.00,0.00,1.00}{Georgia}}} \in \mathbb{A}_{\text{\textcolor[rgb]{0.00,0.00,1.00}{Georgia}}}$ and $a_{\text{\textcolor[rgb]{0.00,0.00,1.00}{Hank}}} > \bar{a}_{\text{\textcolor[rgb]{0.00,0.00,1.00}{Hank}}} \in \mathbb{A}_{\text{\textcolor[rgb]{0.00,0.00,1.00}{Hank}}}$. Here, we conclude the the concave game $\left\langle {\mathbb{I}, {\left( {{\mathbb{A}_i}} \right)}_{i \in \mathbb{I}} ,{\left( {{u_i}} \right)}_{i \in \mathbb{I}}} \right\rangle$ also satisfies the dominance solvability condition in  \cite{moulin1984dominance}. Hence, it satisfies Rosen's  well-known conditions for existence and uniqueness of a pure strategy Nash equilibrium \cite{rosen1965existence}.

\section*{Appendix E: Proof of Corollary \ref{corfn}}
\vspace{-0.00cm}
Substituting $\gamma = 0$, $h = 1$ and $g = 0$ into \eqref{eq:defdef} with \eqref{eq:nd4} and \eqref{eq:nd6} yields $q({0,1}) \ge 0$ and $m({0,1}) \ge 0$. It follows from \eqref{rep1} that $m(g,h) \ge 0$ for $g, h \in [0,1]$. Then, from  \eqref{eq:nda1} we obtain the optimal solution of \eqref{eq:aie1} as ${g^*}\left( h \right) = 0$. By \eqref{rep2}, we obtain from $q({0,1}) \ge 0$ that $q({0,h}) \ge 0$, and from \eqref{eq:nda2}  we have $\frac{{\partial f\!\left( {0,h} \right)}}{{\partial h}} \ge 0$. Thus, the optimal solution of \eqref{eq:aie2} is ${h^*} = 1$. Consequently, the max-min strategy of \eqref{eq:sdwb} is $(g^*, h^*) = (0,1)$, which is also the pure Nash equilibrium via considering Theorem \ref{exth}.

\vspace{-0.10cm}
\section*{Appendix F: Proof of Theorem~\ref{thm:ai}}
As Theorem \ref{exth} states, the pure Nash equilibrium exists for the games \eqref{eq:sdwb} and \eqref{eq:iae}. Therefore, to derive it, we only need to study min-max or max-min strategy. In this proof, we study the max-min problem \eqref{eq:sdwb}.

Based on \eqref{desg} and \eqref{desh}, the rest of proof considers five different cases with the following auxiliary function (assuming $\gamma \ne 0$):
\begin{align}
&\delta(g)  \!\triangleq\!  - \frac{\sqrt {(\beta + 2g\gamma)(\beta + \lambda - 1) - (\lambda + 2\beta - 2)\widehat{s}\gamma - \chi\gamma } }{\gamma} \nonumber\\
&\hspace{1.1cm}- \frac{{1 \!-\! \lambda}}{\gamma} + h. \label{delta}
\end{align}

\vspace{-0.30cm}
\subsection*{Case A: $m(0,1) \ge 0$}
Due to \eqref{rep1},  $m(0,1) \ge 0$, in conjunction with \eqref{eq:nda1},  indicates that given any $h \in [\overline{s},1]$, we have
\begin{align}
\frac{{\partial f\!\left( {g,h} \right)}}{{\partial g}} \ge 0, ~\text{for}~\text{any}~g \in [ {0,{\underline{s}}} ]\label{kvxx}
\end{align}
which implies that  $f\!\left( {g,h} \right)$ is non-decreasing with respect to $g$. Thus, from  \eqref{eq:aie1} we have
\begin{align}
{g^*}\!\left( h \right) = 0. \label{eq:nda7}
\end{align}

We next insert (\ref{eq:nda7}) into (\ref{eopo}) and take its derivative w.r.t. $h$:
\begin{align}
\frac{{\mathrm{d} f( {{g^*}( h),h})}}{{\mathrm{d} h}} = \frac{q(0, h)c^\top\bf{1}}{(1 - \lambda  - \gamma h)^2},\label{eq:A1}
\end{align}
where $q(0,h)$ is given in \eqref{eq:th1}. We note that \eqref{dvh2} indicates that $q(0,h)$ is non-increasing w.r.t. $h$. Thus, if $q({0,1}) \ge 0$, $q({0,h}) \ge 0$ for any $h \in [\overline{s}, 1]$. We conclude from (\ref{eq:A1}) that $f\left( {{g^*}\!\left( h \right),h} \right)$ is non-decreasing w.r.t. $h$. Hence via \eqref{eq:aie2}, $h^* = 1$. If $q({0,{\overline{s}}}) \le 0$, we have $q({0,h}) \le 0$ for any $h \in [\overline{s}, 1]$. Thus, $f\left( {{g^*}\!\left( h \right),h} \right)$ is non-increasing w.r.t.  $h$, and hence $h^* = \overline{s}$.
If $q({0,{\overline{s}}}) > 0$ and $q({0,1}) < 0$, it can be verified from \eqref{eq:defeq2} and (\ref{eq:th1}) that $q({0,r(0)}) = 0$. Then, from (\ref{eq:A1}) we have $\frac{{\mathrm{d} f( {{g^*}( h ),h} )}}{{\mathrm{d} h}} \geq 0$ for $h \in [\overline{s}, r(0)]$ and $\frac{{\mathrm{d} f\!\left( {{g^*}\!\left( h \right),h} \right)}}{{\mathrm{d} h}} < 0$ for $h \in [r(0), 1]$, which implies that $h^* = r(0)$. The equilibrium point in this case is summarized in \eqref{ma1}.

\vspace{-0.20cm}
\subsection*{Case B: $m(\underline{s},\overline{s}) \le 0$}
Due to \eqref{desg} and \eqref{desh}, $m(\underline{s},\overline{s}) \le 0$ implies that given any $h \in [\overline{s},1]$, $\frac{{\partial f(g,h)}}{{\partial g}} \le 0$ for $g \in (0,{\underline{s}}]$, from which and (\ref{eq:aie1}), we have $g^*(h) = \underline{s}$. We now plug $g^*(h)$ into (\ref{eopo}) and take its derivative w.r.t. $h$:\begin{align}
\frac{{\mathrm{d} f(g^*(h), h)}}{{\mathrm{d} h}} = \frac{{q(\underline{s}, h )c^ \top\bf{1}}}{{\left( {1 - \lambda  - \gamma h + \gamma} \right)}^2}.\label{eq:A1cs2}
\end{align}

Due to  \eqref{dvh2}, if $q({\underline{s},1}) \ge 0$, $q({\underline{s},h}) \ge 0$ for any $h \in [\overline{s}, 1]$. We conclude from (\ref{eq:A1cs2}) that $f( {\underline{s},h})$ is non-decreasing w.r.t $h$; thus, $h^* = 1$. If $q(\underline{s},\overline{s}) \le 0$, then $q(\underline{s},h) \le 0$ for any $h \in [\overline{s}, 1]$. Thus, $f(g^*( h ),h)$ is non-increasing w.r.t $h$, and hence $h^* = \overline{s}$. If $q(\underline{s},\overline{s}) > 0$ and $q(\underline{s},1) < 0$, it can be verified  from \eqref{eq:defeq2} and \eqref{eq:th1} that $q({\underline{s},r(\underline{s})}) = 0$. Then, via (\ref{eq:A1cs2}), we have $\frac{{\mathrm{d} f(\underline{s},h)}}{{\mathrm{d} h}} \geq 0$ for $h \in [\overline{s}, r(\underline{s})]$ and $\frac{{\mathrm{d} f(\underline{s},h)}}{{\mathrm{d} h}} < 0$ for $h \in [r(\underline{s}), 1]$, which implies that $h^* = r(\underline{s})$.
The equilibrium  is summarized as
\begin{align}
\text{If}~m(\underline{s},\overline{s}) \le 0,      ~~(g^*,h^*) \!=\! \begin{cases}
		(\underline{s},1), &\!\!\text{if } q(\underline{s},1) \!\ge\! 0\\
		(\underline{s},\overline{s}), &\!\!\text{if } q(\underline{s},\overline{s}) \!\le\! 0\\
		(\underline{s},r(\underline{s})), &\!\!\text{otherwise}.
	\end{cases}\label{maad}
  \end{align}

By Lemma \ref{ladd1}, the condition $m(\underline{s},\overline{s}) \le 0$ \& $q(\underline{s},\overline{s}) \le 0$ in \eqref{maad} contradicts with $m(\underline{s},\overline{s}) + q(\underline{s},\overline{s}) > 0$. The ``otherwise" in \eqref{maad} represents $q(\underline{s},1) < 0$ \& $q(\underline{s},\overline{s}) > 0$ \& $m(\underline{s},\overline{s}) \le 0$, which with \eqref{desh} imply
$q(\underline{s},1) + m(\underline{s},1) \le q(\underline{s},1) + m(\underline{s},\overline{s}) < 0$.  Note that this contradicts with $q(\underline{s},1) + m(\underline{s},1) \ge 0$, which  is a consequence of Lemma \ref{ladd1}. Thus, the ``otherwise" condition in \eqref{maad} does not hold as well. Therefore, we have:
\begin{align}
\text{If}~m(\underline{s},\overline{s}) \le 0, ~~(g^*,h^*) = (\underline{s},1).\label{ma2x}
\end{align}

\vspace{-0.50cm}
\subsection*{Case C: $m(0,\overline{s}) \le 0$ \& $m(\underline{s},1) \ge 0$}
It follows from \eqref{desh} that $m(0,\overline{s}) \le 0$ and $m(\underline{s},1) \ge 0$ imply $m(0,h) \le 0$ and $m(\underline{s},h) \ge 0$ for $h \in [\overline{s},1]$, which indicate that $\frac{{\partial f\left( {g,h} \right)}}{{\partial g}} \ge 0$ for any $g \in (w(h),\underline{s}]$, and $\frac{{\partial f\left( {g,h} \right)}}{{\partial g}} \le 0$ for any $g \in [0, w(h)]$, where $w(h)$ is given by \eqref{dtm}. The relation ${g^*}(h)=w(h)$  follows from \eqref{eq:aie1}, whose derivate w.r.t. $h$ is:
\begin{align}
\frac{{\mathrm{d} {g^*}(h)}}{{\mathrm{d} h}} =  \frac{{\mathrm{d} w(h)}}{{\mathrm{d} h}}
 \!=\! 1 \!-\! \frac{1 \!-\! \lambda  \!-\! \beta }{\Xi}, \label{kg1}
\end{align}
where
\begin{equation}
	\Xi\!=\!\sqrt {(1 \!-\! \lambda )(1 \!-\! \lambda  \!-\! 2h\gamma  \!-\! \beta ) \!+\! (2 \!-\! \lambda  \!-\! 2\beta )\gamma \widehat s \!+\! 2h\beta \gamma  \!-\! \chi\gamma }\nonumber
\end{equation}
Replacing $g$ in \eqref{eopo} by ${g^*}(h)$ and taking its derivative w.r.t. $h$, we get
\begin{align}
\frac{{\mathrm{d} f( {{g^*}( h ),h})}}{{\mathrm{d} h}} = \frac{{( {{r_1}{r_3} \!+\! {r_2} \!+\! ( {{r_1}{r_4} \!-\! {r_2}} )\frac{{\mathrm{d} {g^*}(h)}}{{\mathrm{d} h}}}){{c}^ \top }{{\bf{1}}}}}{{{{( {{a_2} \!+\! g\gamma })}^2}}}, \label{kg2}
\end{align}
where we define:
\begin{subequations}
\begin{align}
{r_1} &\triangleq  1 - \lambda  - h\gamma  + \gamma {g^*}( h ),\\
{r_2} &\triangleq  ( {1 - 2\beta  + h\gamma } )\widehat s\gamma  + h\beta \gamma  - {h^2}{\gamma ^2} - \chi\gamma \nonumber\\
 &\hspace{2.95cm}+ ( {\beta  \!-\! \gamma \widehat s \!+\! \gamma {g^*}(h)})\gamma {g^*}(h),\\
{r_3} &\triangleq \beta  + \widehat s\gamma  - 2h\gamma ,\\
{r_4} &\triangleq \beta  - \widehat s\gamma  + 2\gamma {g^*}( h ).
\end{align}\label{kg3}
\end{subequations}
Since $g^*(h) = w(h) \ge 0$ due to \eqref{dtm}, we have
\begin{align}
\Xi\ge 1 - \lambda  - h\gamma \ge 0. \label{kg5}
\end{align}
Moreover, $1 - \lambda  - h\gamma  - (1 - \lambda  - \beta ) = \beta  - h\gamma  \ge 0$, which implies that $1 - \lambda  - h\gamma  \ge 1 - \lambda  - \beta $. Then, noting \eqref{kg5} and \eqref{eq:a2con}, we obtain
$\Xi  \ge 1 - \lambda  - h\gamma \ge 1 - \lambda  - \beta \ge 0$,
which, in conjunction with \eqref{kg1}, yields
\begin{align}
0 \le \frac{{\mathrm{d} {{g^*}( h )}}}{{\mathrm{d} h}} \le 1\label{kg6}.
\end{align}

If ${r_1}{r_4} - {r_2} \le 0$, it follows from \eqref{kg3} and \eqref{kg6} that
\begin{align}
&{r_1}{r_3} + {r_2} + ({r_1}{r_4} - {r_2})\frac{{\mathrm{d} {g^*}(h)}}{{\mathrm{d} h}} \nonumber\\
& \ge {r_1}{r_3} + {r_2} + {r_1}{r_4} - {r_2}\label{kg777}\\
& = 2\left( {1 - \lambda  - h\gamma  + \gamma {g^*}(h)} \right)\left( {\beta  - h\gamma  + \gamma {g^*}(h)} \right) \ge 0,\label{kg7}
\end{align}
where \eqref{kg777} follows from \eqref{eq:a2con} and \eqref{eq:asfx00}.

From \eqref{mean}, we have:
\begin{align}
\chi \le \frac{1}{{\sum\limits_{i \in \mathbb{V}} {{c_{i}}} }}\sum\limits_{i \in \mathbb{V}} {{{c}_i}{s_i}} w_{\max}
= w_{\max}\widehat{s}. \label{kg7kk}
\end{align}
where we use  $w_{\max}\triangleq\max \limits_{i \in \mathbb{V}} \sum \limits_{j \in \mathbb{V}} w_{ij}$. We obtain from \eqref{kg3} that
\begin{align}
&( {{r_1}{r_3} + {r_2}} ) - ({r_1}{r_4} - {r_2})\nonumber\\
& = 2\gamma {( {{g^*}(h)})^2} - 2\gamma {h^2} + 2h\gamma \widehat s - 2g^*(h)\gamma \widehat s + h\beta  + {g^*}(h)\beta  \nonumber\\
&\hspace{0.5cm}- \chi - 2\beta \widehat s + h + {g^*}(h) + \lambda \widehat s - \lambda h - \lambda {g^*}(h)\label{kg8a1}\\
& \ge  - 2\gamma {h^2} + 2h\gamma \widehat s + h\beta  - \chi - 2\beta \widehat s + h + \lambda \widehat s - \lambda h \label{kg8a2}\\
& \ge  - \widehat s\beta  - \chi + \widehat s \label{kg8a3}\\
&\ge \left( {1 - {w_{\max }} - \beta } \right)\widehat{s} > 0, \label{kg8}
\end{align}
where \eqref{kg8a2} follows from \eqref{kg8a1} considering $(1+\beta-\lambda-2\gamma\widehat{s} + 2\gamma g^*(h)) g^*(h) \ge 0$ (due to \eqref{eq:a2con}), and   \eqref{kg8a3} follows from \eqref{kg8a2} due to $-2h^2\gamma + (2\gamma\widehat{s} + 1 + \beta -\lambda)h \ge (1+\beta-\lambda)\widehat{s}$, since $\frac{1+\beta - \lambda + 2\gamma\widehat{s}}{4\gamma} \ge 1$ and $h \in [\widehat{s},1]$. We note also that \eqref{kg8} follows from \eqref{kg8a3} due to \eqref{kg7kk}.

If ${r_1}{r_4} - {r_2} > 0$, from \eqref{kg8} we have ${r_1}{r_3} + {r_2}$ $\ge$ ${r_1}{r_4} - {r_2} > 0$. From \eqref{kg6}, we have  ${r_1}{r_3} + {r_2} + ({r_1}{r_4} - {r_2})\frac{{\mathrm{d} {g^*}(h)}}{{\mathrm{d} h}} \ge {r_1}{r_3} + {r_2} > 0$, which, in conjunction with \eqref{kg7} and \eqref{kg2}, results in $\frac{{\mathrm{d} f( {{g^*}( h ),h})}}{{\mathrm{d} h}} \ge 0$. We obtain here $h^* = 1$, and consequently, $g^* = w(1)$.  The equilibrium in this case is expressed as:
\begin{align}
\text{If}~m(0,\overline{s}) \le 0 ~\&~ m(\underline{s},1) \ge 0, ~~(g^*,h^*) = (w(1),1).\label{kg9}
\end{align}

\vspace{-0.50cm}
\subsection*{Case D: $m({\underline{s},{\overline{s}}}) > 0$ \& $m({{0},1}) < 0$ \& $m({0,{\overline{s}}}) > 0$)}
Due to \eqref{desh}, we obtain from $m({{0},1}) < 0$ \& $m({0,{\overline{s}}}) > 0$:
\begin{subequations}
\begin{align}
m(0, h) &\le 0, h \in [\delta(0), 1], \label{cd1a}\\
m(0, h) &> 0, h \in [\overline{s}, \delta(0)).\label{cd1b}
\end{align}\label{cd1}
\end{subequations}
\noindent It follows from \eqref{desg} and \eqref{cd1b}  that $m(g, h) > 0$ for $h \in [\overline{s}, \delta(0))$ and $g \in [0,\underline{s}]$. Thus, we have $g^* = 0$. Then, following the same analysis in Case A, we arrive at
\begin{align}
\!\!\!\text{For}~h \!\in\! [\overline{s}, \delta(0)), (g^*,h^*) \!\!=\!\! \begin{cases}
		\!(0,\delta(0)), &\!\!\!\text{if } q(0,\delta(0)) \!\ge\! 0\\
		\!(0,\overline{s}), &\!\!\! \text{if } q(0,\overline{s}) \!\le\! 0\\
		\!(0,r(0)), &\!\!\!\text{otherwise}.
\end{cases}\label{cd3}
\end{align}
We note that the ``otherwise" in \eqref{cd3}  is $q(0,\delta(0)) < 0$ \& $q(0,\overline{s}) > 0$. Following \eqref{dvh2}, we have $q(0,1) \le q(0,\overline{s}) \le 0$ or $q(0,1) \le q(0,\delta(0)) \le 0$. Noticing $m({{0},1}) < 0$, we obtain $q(0,1) + m(0,1)  < 0$ that contradicts with $q(0,1) + m(0,1)  > 0$ implied by Lemma \ref{ladd1}. Thus, the conditions of the second and third items in \eqref{cd3} do not hold. Therefore, \eqref{cd3} in this case can be expressed as
\begin{align}
\text{For}~h \in [\overline{s}, \delta(0)), ~~(g^*,h^*) = (0,\delta(0)).\label{cd3ff}
\end{align}

If $m(\underline{s}, 1) \ge 0$, we have  $m(\underline{s}, h) \ge 0$ for $h \in [\overline{s},1]$, which follows from \eqref{desh}. With the consideration of \eqref{cd1a}, following the same analysis in Case C, we obtain
\begin{align}
\emph{\emph{For}}~h \!\in\! [\delta(0),1)~\&~m(\underline{s}, 1) \!\ge\! 0, ~(g^*,h^*) \!=\! (w(1), 1).\label{cd4}
\end{align}

Since \eqref{cd3ff} and \eqref{cd4} are, respectively, based on $m(0,\delta(0)) = 0$ and $q(w(1), 1) = 0$, due to Lemma \ref{kkm1} we have $f(w(1), 1) \ge f(0,\delta(0))$. From \eqref{cd3ff} and \eqref{cd4}, we obtain
\begin{align}
&\text{If}~m({\underline{s},{\overline{s}}}) \!>\! 0 ~\&~ m({{0},1}) \!<\! 0 ~\&~ m({0,{\overline{s}}}) \!>\! 0~\&~m(\underline{s}, 1) \!\ge\! 0,\nonumber\\
&\hspace{4.7cm} (g^*,h^*) = (w(1),1).\label{ref1}
\end{align}

If $m(\underline{s}, 1) < 0$, we have $m(g, 1) < 0$ for $g\in [0, \underline{s}]$, which follows from \eqref{desg}. We note that $m(0, \overline{s}) > 0$ implies that $m(g, \overline{s}) > 0$ for $g\in [0, \underline{s}]$. Noting \eqref{desh}, we conclude:
\begin{subequations}
\begin{align}
m(g, h) &> 0~ \text{for}~\text{any}~h \in [\overline{s}, \delta(g)], \label{cd5a}\\
m(g, h) &\le 0~ \text{for}~\text{any}~ h \in (\delta(g), 1],\label{cd5b}
\end{align}\label{cd5}
\end{subequations}
\noindent where $\delta(g)$ is given in \eqref{delta}. Taking its derivative w.r.t. $g$, we have:
\begin{align}
\frac{\mathrm{d} \delta(g)}{\mathrm{d} g} = & \frac{1 \!-\! \lambda \!-\! \beta}{\sqrt {(\beta \!+\! 2g\gamma)(\beta \!+\! \lambda \!-\! 1) \!-\! (\lambda \!+\! 2\beta \!-\! 2)\widehat{s}\gamma \!-\! \bar w\gamma }} \nonumber\\
&\hspace{0.1cm} +  1 > 0, \label{kb}
\end{align}
where the inequality is obtained via considering \eqref{eq:a2con}. Since \eqref{kb} implies that $\delta(g)$ is an increasing function,  we have $\delta(\underline{s}) \ge \delta(0)$. It follows from \eqref{eq:defdef} with \eqref{delta} that $m(0, \delta(0)) = 0$ and $m(\underline{s}, \delta(\underline{s})) = 0$, which respectively imply $m(0, h) \le 0$ and $m(\underline{s},h) \ge 0$ for $h \in [\delta(0), \delta(\underline{s})]$ (that is due to \eqref{desh}). Following the analysis in Case C, we obtain
  \begin{align}
&\emph{\emph{For}}~h \!\in\! [\delta(0), \delta(\underline{s})]~\&~ m(\underline{s},1) < 0,  \nonumber\\
&\hspace{3.8cm}  (g^*,h^*) \!=\! (w(\delta(\underline{s})), \delta(\underline{s})). \label{cd7}
  \end{align}

Noting \eqref{eq:nda1}, we obtain from \eqref{cd5b} that $g^* = \underline{s}$ for $h \in [\delta(\underline{s}), 1]$. Following the analysis in Case B, we arrive at
  \begin{align}
&\emph{\emph{For}}~h \!\in\! [\delta(\underline{s}), 1]~\&~ m(\underline{s},1) < 0,  \nonumber\\
&\hspace{1.8cm}  (g^*,h^*) \!=\! \begin{cases}
		(\underline{s},1), & \text{if } q(\underline{s},1) \ge 0\\
		(\underline{s},\delta(\underline{s})), & \text{if } q(\underline{s},\delta(\underline{s})) \le 0\\
		(\underline{s},r(\underline{s})), & \text{otherwise}.
	\end{cases}\label{cd8}
  \end{align}

We note that ``otherwise" condition in  \eqref{cd8} is $q(\underline{s},1) < 0$ \& $q(\underline{s},\delta(\underline{s})) > 0$. Due to \eqref{dvh1}, we have $q(\underline{s},1) \le q(\underline{s},\delta(\underline{s})) \le 0$,   which, in conjunction with $m(\underline{s},1) < 0$, results in $q(\underline{s},1) + m(\underline{s},1) < 0$ that contradicts with $q(\underline{s},1) + m(\underline{s},1) > 0$ implied by  Lemma \ref{ladd1}. Thus, the conditions of the second and the third items in \eqref{cd8} do not hold. Thus, \eqref{cd8} is simplified as
\begin{align}
&\text{For}~h \in[\delta(\underline{s}), 1]~\&~ m(\underline{s},1) < 0~\&~q(\underline{s},1) \ge 0,  \nonumber\\
&\hspace{5.2cm}(g^*,h^*) = (\underline{s},1).\label{cd8ff}
  \end{align}

Noting $w(\delta(\underline{s})) \le \underline{s}$, $w(1) \le \underline{s}$, the condition $q(\underline{s},1) \ge 0$ in \eqref{cd8ff}, the result in \eqref{cd7} is based on $m(w(\delta(\underline{s})), \delta(\underline{s})) = 0$, and
the result in \eqref{cd3ff} is based on $m(0, \delta(0)) \ge 0$. By Lemma \ref{kkm1} we obtain $f(\underline{s},1) \ge f(w(\delta(\underline{s})),\delta(\underline{s}))$ and $f(\underline{s},1) \ge f(0, \delta(0))$, which, in conjunction with \eqref{cd3ff}, \eqref{cd7} and \eqref{cd8ff}, yields the equilibrium:
\begin{align}
&\text{If}~m({\underline{s},{\overline{s}}}) \!>\! 0 ~\&~ m({{0},1}) \!<\! 0 ~\&~ m({0,{\overline{s}}}) \!>\! 0~\&~m(\underline{s}, 1) \!<\! 0,\nonumber\\
&\hspace{5.4cm} (g^*,h^*) = (\underline{s},1).\label{ref2}
\end{align}

\vspace{-0.50cm}
\subsection*{Case E: $m({\underline{s},{\overline{s}}}) > 0$ \& $m({{0},1}) < 0$ \& $m({\underline{s},1}) < 0$)}
Noting \eqref{desg}, we obtain from $m({\underline{s},{\overline{s}}}) > 0$ \& $m({\underline{s},1}) < 0$ that $m(\underline{s},\delta(\underline{s})) = 0$,
where $\delta(\underline{s})$ is given by \eqref{delta}. Consequently,
\begin{subequations}
\begin{align}
m(\underline{s}, h) &\le 0, h \in [\delta(\underline{s}), 1], \label{ce1a}\\
m(\underline{s}, h) &> 0, h \in [\overline{s}, \delta(\underline{s})).\label{ce1b}
\end{align}\label{ce1}
\end{subequations}
It follows from \eqref{desg} and \eqref{ce1a}  that $m(g, h) \le 0$ for $h \in [\delta(\underline{s}), 1]$ and $g \in [0,\underline{s}]$. Thus, we have $g^* = \underline{s}$. Then, following the analysis in Case B, we arrive at
\begin{align}
\!\!\!\text{For}~\!h \!\in\! [\delta(\underline{s}), 1], (g^*,h^*) \!=\! \begin{cases}
		\!(\underline{s},1), &\!\!\text{if } q(\underline{s},1) \!\ge\! 0\\
		\!(\underline{s},\delta(\underline{s})), &\!\!\text{if } q(\underline{s},\delta(\underline{s})) \!\le\! 0\\
		\!(\underline{s},r(\underline{s})), &\!\!\text{otherwise}.
	\end{cases}\label{vva1}
  \end{align}

We note that ``otherwise" in \eqref{vva1} includes the condition  $q(\underline{s},1) < 0$. By \eqref{dvh2}, we have $q(\underline{s},1) \le q(\underline{s},\delta(\underline{s})) \le 0$. Noticing $m(\underline{s},1) < 0$ in this case, we have $q(\underline{s},1) + m(\underline{s},1)  < 0$, which contradicts with
$q(\underline{s},1) + m(\underline{s},1)  > 0$ implied by Lemma \ref{ladd1}. Thus, the conditions of the second and third items in \eqref{vva1} do not hold. Therefore,  \eqref{vva1}  reduces to
\begin{align}
\!\!\!\text{For}~\!h \!\in\! [\delta(\underline{s}), 1] ~\&~ q(\underline{s},1) \ge 0, ~~(g^*,h^*) = (\underline{s},1). \label{ce2}
  \end{align}

If $m({0,{\overline{s}}}) \le 0 $, from \eqref{desh} we have  $m(0, h) \le 0$ for $h \in [\overline{s},1]$. With the consideration of \eqref{ce1b}, following the analysis in Case C, we obtain
\begin{align}
&\text{For}~h \in [\overline{s}, \delta(\underline{s}))~\&~m(0,\overline{s}) \le 0, \nonumber\\
&\hspace{3.8cm}(g^*,h^*) = (w(\delta(\underline{s})), \delta(\underline{s})).\label{ce4}
  \end{align}

Noting $1 \geq \delta(\underline{s})$, $\underline{s} \ge w(\delta(\underline{s}))$,  the result in \eqref{ce4} is based on $m(w(\delta(\underline{s})), \delta(\underline{s})) = 0$ and the condition $q(\underline{s},1) \ge 0$ in \eqref{ce2}. From Lemma \ref{kkm1} we have $f(\underline{s},1) \ge f(w(\delta(\underline{s})),\delta(\underline{s}))$. Consequently, we get
\begin{align}
&\text{If}~m({\underline{s},{\overline{s}}}) \!>\! 0 ~\&~ m({{0},1}) \!<\! 0 ~\&~ m({\underline{s},1}) \!<\! 0 ~\&~m(0,\overline{s}) \!\le\! 0,\nonumber\\
&\hspace{5.2cm} (g^*,h^*) = (\underline{s},1).\label{ref3}
\end{align}

If $m({0,{\overline{s}}}) > 0$, we have $m(g, \overline{s}) > 0$ for $g\in [0, \underline{s}]$, which is due to \eqref{desg}. We note that $m({\underline{s},1}) < 0$ implies that $m(g, 1) < 0$ for $g\in [0, \underline{s}]$. Here, we  conclude \eqref{cd5}. Considering \eqref{eq:nda1}, we obtain from \eqref{cd5a} that $g^* = 0$ for $h \in [\overline{s}, \delta(0)]$. Following the analysis in Case A, we have  \begin{align}
&\emph{\emph{For}}~h \!\in\! [\overline{s}, \delta(0)]~\&~ m({0,{\overline{s}}}) > 0,  \nonumber\\
&\hspace{1.7cm}  (g^*,h^*) = \begin{cases}
		(0,\delta(0)), & \text{if } q(0,\delta(0)) \ge 0\\
		(0,\overline{s}), & \text{if } q(0,\overline{s}) \le 0\\
		(0,r(0)), & \text{otherwise}.
	\end{cases}\label{vva2}
  \end{align}

The ``otherwise" in \eqref{vva2} includes the condition $q(0,\delta(0)) < 0$. Due to \eqref{dvh2}, we have $q(0,1) \le q(0,\overline{s}) \le 0$ and $q(0,1) \le q(0,\delta(0)) < 0$. Noting $m(0,1) < 0$ in this case, we have $m(0,1) + q(0,1)  < 0$, which contradicts with
$q(0,1) + m(0,1)  > 0$ implied by Lemma \ref{ladd1}. Thus, the conditions of the second the third items in \eqref{vva2} do not hold. Therefore,  \eqref{vva2}  reduces to
\begin{align}
&\emph{\emph{For}}~h \!\in\! [\overline{s}, \delta(0)]~\&~ m({0,{\overline{s}}}) > 0~\&~q(0,\delta(0)) \ge 0, \nonumber \\ &\hspace{4.5cm} (g^*,h^*) = (0,\delta(0)). \label{ce8}
  \end{align}
 Following the steps used in the derivation of \eqref{cd7}, we obtain
  \begin{align}
&\text{For}~h \!\in\! [\delta(0), \delta(\underline{s})]~\&~ m({0,{\overline{s}}}) > 0,  \nonumber\\
&\hspace{3.8cm}  (g^*,h^*) \!=\! (w(\delta(\underline{s})), \delta(\underline{s})). \label{ce7}
  \end{align}
We note that \eqref{vva2} (which leads to \eqref{ce8}) is based on the condition $m(g, h) > 0$ for $h \in [\overline{s}, \delta(0))$ and $g \in [0,\underline{s}]$, which implies $m(0,\delta(0)) > 0$. Moreover, for \eqref{ce7}, we have $m(w(\delta(\underline{s})), \delta(\underline{s})) = 0$.
Then, by Lemma \ref{kkm1}, from \eqref{ce2},  \eqref{ce8} and \eqref{ce7} we have $f(\underline{s},1) \ge f(0,\delta(0))$ and $f(\underline{s},1) \ge f(w(\delta(\underline{s})),\delta(\underline{s}))$, and combining the conditions in \eqref{ce2},  \eqref{ce8} and \eqref{ce7}, we arrive at
\begin{align}
&\text{If}~m({\underline{s},{\overline{s}}}) \!>\! 0 ~\&~ m({{0},1}) \!<\! 0 ~\&~ m({\underline{s},1}) \!<\! 0 ~\&~m(0,\overline{s}) \!>\! 0,\nonumber\\
&\hspace{5.2cm} (g^*,h^*) = (\underline{s},1).\label{ref4}
\end{align}

\vspace{-0.70cm}
\subsection*{Summary for Cases B--E}
We note that \eqref{ma1} is obtained in Case A. Combining \eqref{ref1} and \eqref{kg9}  yields \eqref{ma3}, while combining \eqref{ma2x}, \eqref{ref2}, \eqref{ref3} and \eqref{ref4} results in \eqref{ma2}.

\section*{Appendix G: Proof of Theorem \ref{influence}}
It follows from \eqref{eq:defdef} with \eqref{eq:nd4} and \eqref{eq:nd6} that
\begin{align}
&m(0,1) = ( {1 \!-\! \lambda  \!-\! 2\gamma  \!+\! 2\widehat{s}\gamma })\beta  - ( {2\widehat{s} \!-\! \lambda \widehat{s} \!-\! \gamma  \!-\! \chi })\gamma. \label{pqf2}
\end{align}

We note that \eqref{eq:hhk} implies $1 - \lambda  - 2\gamma  + 2\widehat{s}\gamma > 0$, which implies that if we require $m(0,1) < 0$, we must have ${2\widehat{s} - \lambda \widehat{s} - \gamma  - \chi } > 0$, which means the left-hand of \eqref{bob}. Then noticing $0 \le \frac{\gamma}{\beta} \le 1$, but $ \frac{\gamma}{\beta} = 0$ implies $\gamma = 0$, we straightforwardly verify from \eqref{pqf2} that \eqref{bob} is equivalent to $m(0,1) < 0$. Following the same analysis, we conclude that \eqref{Alice} is equivalent to $q(0,1) < 0$.

From \eqref{ma1}--\eqref{ma3} we  conclude that the CB influences \textcolor[rgb]{0.00,0.00,1.00}{Georgia}'s  strategy (i.e.,  $g^* \neq 0$) if and only if $m(0,1) < 0$. Thus, the proof of \eqref{bob} is established.

From \eqref{ma1}  we conclude that the CB influences \textcolor[rgb]{0.00,0.00,1.00}{Hank's} strategy (i.e., $h^* \neq 1$) if and only if $m(0,1) \ge 0$ and $q(0,1) < 0$. We note that $q(0,1) < 0$ and $m(0,1) < 0$ simultaneously do not hold, since the condition $q(0,1) + m(0,1) > 0$ implied by Lemma \ref{ladd1} is violated. Therefore, the sufficient and necessary condition is   $q(0,1) < 0$.

\bibliographystyle{IEEEtran}
\bibliography{ref}

\begin{thebibliography}{10}
\providecommand{\url}[1]{#1}
\csname url@samestyle\endcsname
\providecommand{\newblock}{\relax}
\providecommand{\bibinfo}[2]{#2}
\providecommand{\BIBentrySTDinterwordspacing}{\spaceskip=0pt\relax}
\providecommand{\BIBentryALTinterwordstretchfactor}{4}
\providecommand{\BIBentryALTinterwordspacing}{\spaceskip=\fontdimen2\font plus
\BIBentryALTinterwordstretchfactor\fontdimen3\font minus
  \fontdimen4\font\relax}
\providecommand{\BIBforeignlanguage}[2]{{%
\expandafter\ifx\csname l@#1\endcsname\relax
\typeout{** WARNING: IEEEtran.bst: No hyphenation pattern has been}%
\typeout{** loaded for the language `#1'. Using the pattern for}%
\typeout{** the default language instead.}%
\else
\language=\csname l@#1\endcsname
\fi
#2}}
\providecommand{\BIBdecl}{\relax}
\BIBdecl

\bibitem{asilomar}
Y.~Mao and E.~Akyol, ``Competitive information spread with confirmation bias,''
  in \emph{53rd Asilomar Conference on Signals, Systems, and Computers}, pp.
  391--395, 2019.

\bibitem{moreno1934shall}
J.~L. Moreno, ``Who shall survive?: A new approach to the problem of human
  interrelations.'' 1934.

\bibitem{french1956formal}
J.~R. French~Jr, ``A formal theory of social power.'' \emph{Psychological
  review}, vol.~63, no.~3, p. 181, 1956.

\bibitem{degroot1974reaching}
M.~H. DeGroot, ``Reaching a consensus,'' \emph{Journal of the American
  Statistical Association}, vol.~69, no. 345, pp. 118--121, 1974.

\bibitem{abelson1964mathematical}
R.~P. Abelson, ``Mathematical models of the distribution of attitudes under
  controversy,'' \emph{Contributions to mathematical psychology}, 1964.

\bibitem{friedkin1990social}
N.~E. Friedkin and E.~C. Johnsen, ``Social influence and opinions,''
  \emph{Journal of Mathematical Sociology}, vol.~15, no. 3-4, pp. 193--206,
  1990.

\bibitem{dhamal2018optimal}
S.~Dhamal, W.~Ben-Ameur, T.~Chahed, and E.~Altman, ``Optimal investment
  strategies for competing camps in a social network: A broad framework,''
  \emph{IEEE Transactions on Network Science and Engineering, DOI:
  10.1109/TNSE.2018.2864575.}

\bibitem{das2013debiasing}
A.~Das, S.~Gollapudi, R.~Panigrahy, and M.~Salek, ``Debiasing social wisdom,''
  in \emph{Proceedings of the 19th ACM SIGKDD international conference on
  Knowledge discovery and data mining}, 2013, pp. 500--508.

\bibitem{hegselmann2002opinion}
R.~Hegselmann and U.~Krause, ``Opinion dynamics and bounded confidence models,
  analysis, and simulation,'' \emph{Journal of artificial societies and social
  simulation}, vol.~5, no.~3, 2002.

\bibitem{proskurnikov2017tutorial}
A.~V. Proskurnikov and R.~Tempo, ``A tutorial on modeling and analysis of
  dynamic social networks. {Part I},'' \emph{Annual Reviews in Control},
  vol.~43, pp. 65--79, 2017.

\bibitem{proskurnikov2018tutorial}
------, ``A tutorial on modeling and analysis of dynamic social networks. {Part
  II},'' \emph{Annual Reviews in Control}, vol.~45, pp. 166--190, 2018.

\bibitem{xu2020paradox}
C.~Xu, J.~Li, T.~Abdelzaher, H.~Ji, B.~K. Szymanski, and J.~Dellaverson, ``The
  paradox of information access: On modeling social-media-induced
  polarization,'' \emph{arXiv:2004.01106}.

\bibitem{giridhar2019social}
P.~Giridhar and T.~Abdelzaher, ``Social media signal processing,''
  \emph{Social-Behavioral Modeling for Complex Systems}, pp. 477--493, 2019.

\bibitem{cui2019semi}
H.~Cui, T.~Abdelzaher, and L.~Kaplan, ``A semi-supervised active-learning truth
  estimator for social networks,'' in \emph{The World Wide Web Conference}, pp.
  296--306, 2019.

\bibitem{nickerson1998confirmation}
R.~S. Nickerson, ``Confirmation bias: A ubiquitous phenomenon in many guises,''
  \emph{Review of general psychology}, vol.~2, no.~2, pp. 175--220, 1998.

\bibitem{lazer2018science}
D.~M. Lazer, M.~A. Baum, Y.~Benkler, A.~J. Berinsky, K.~M. Greenhill,
  F.~Menczer, M.~J. Metzger, B.~Nyhan, G.~Pennycook, D.~Rothschild
  \emph{et~al.}, ``The science of fake news,'' \emph{Science}, vol. 359, no.
  6380, pp. 1094--1096, 2018.

\bibitem{vicario2019polarization}
M.~D. Vicario, W.~Quattrociocchi, A.~Scala, and F.~Zollo, ``Polarization and
  fake news: Early warning of potential misinformation targets,'' \emph{ACM
  Transactions on the Web}, vol.~13, no.~2, pp. 1--22, 2019.

\bibitem{kappes2020confirmation}
A.~Kappes, A.~H. Harvey, T.~Lohrenz, P.~R. Montague, and T.~Sharot,
  ``Confirmation bias in the utilization of others’ opinion strength,''
  \emph{Nature Neuroscience}, vol.~23, no.~1, pp. 130--137, 2020.

\bibitem{zhao2016competitiveness}
J.~Zhao, Q.~Liu, L.~Wang, and X.~Wang, ``Competitiveness maximization on
  complex networks,'' \emph{IEEE Transactions on Systems, Man, and Cybernetics:
  Systems}, vol.~48, no.~7, pp. 1054--1064, 2016.

\bibitem{rusinowska2019opinion}
A.~Rusinowska and A.~Taalaibekova, ``Opinion formation and targeting when
  persuaders have extreme and centrist opinions,'' \emph{Journal of
  Mathematical Economics}, vol.~84, pp. 9--27, 2019.

\bibitem{grabisch2018strategic}
M.~Grabisch, A.~Mandel, A.~Rusinowska, and E.~Tanimura, ``Strategic influence
  in social networks,'' \emph{Mathematics of Operations Research}, vol.~43,
  no.~1, pp. 29--50, 2018.

\bibitem{eshghi2018spread}
S.~Eshghi, V.~Preciado, S.~Sarkar, S.~Venkatesh, Q.~Zhao, R.~D'Souza, and
  A.~Swami, ``Spread, then target, and advertise in waves: Optimal budget
  allocation across advertising channels,'' \emph{IEEE Transactions on Network
  Science and Engineering, DOI: 10.1109/TNSE.2018.2873281.}

\bibitem{proskurnikov2016opinion}
A.~V. Proskurnikov, A.~S. Matveev, and M.~Cao, ``Opinion dynamics in social
  networks with hostile camps: {Consensus vs. Polarization},'' \emph{IEEE
  Transactions on Automatic Control}, vol.~61, no.~6, pp. 1524--1536, 2016.

\bibitem{altafini2012consensus}
C.~Altafini, ``Consensus problems on networks with antagonistic interactions,''
  \emph{IEEE Transactions on Automatic Control}, vol.~58, no.~4, pp. 935--946,
  2012.

\bibitem{yang2014opinion}
Y.~Yang, D.~V. Dimarogonas, and X.~Hu, ``Opinion consensus of modified
  hegselmann--krause models,'' \emph{Automatica}, vol.~50, no.~2, pp. 622--627,
  2014.

\bibitem{pineda2013noisy}
M.~Pineda, R.~Toral, and E.~Hern{\'a}ndez-Garc{\'\i}a, ``The noisy
  hegselmann-krause model for opinion dynamics,'' \emph{The European Physical
  Journal B}, vol.~86, no.~12, p. 490, 2013.

\bibitem{mao2018spread}
Y.~Mao, S.~Bouloki, and E.~Akyol, ``Spread of information with confirmation
  bias in cyber-social networks,'' \emph{IEEE Transactions on Network Science
  and Engineering (Special Issue on Network of Cyber-Social Networks: Modeling,
  Analyses, and Control), DOI: 10.1109/TNSE.2018.2878377.}

\bibitem{mas2014cultural}
M.~M{\"a}s, A.~Flache, and J.~A. Kitts, ``Cultural integration and
  differentiation in groups and organizations,'' in \emph{Perspectives on
  culture and agent-based simulations}.\hskip 1em plus 0.5em minus 0.4em\relax
  Springer, 2014, pp. 71--90.

\bibitem{duggins2014psychologically}
P.~Duggins, ``A psychologically-motivated model of opinion change with
  applications to american politics,'' \emph{arXiv preprint arXiv:1406.7770},
  2014.

\bibitem{del2017modeling}
M.~Del~Vicario, A.~Scala, G.~Caldarelli, H.~E. Stanley, and W.~Quattrociocchi,
  ``Modeling confirmation bias and polarization,'' \emph{Scientific reports},
  vol.~7, p. 40391, 2017.

\bibitem{del2016spreading}
M.~Del~Vicario, A.~Bessi, F.~Zollo, F.~Petroni, A.~Scala, G.~Caldarelli, H.~E.
  Stanley, and W.~Quattrociocchi, ``The spreading of misinformation online,''
  \emph{Proceedings of the National Academy of Sciences}, vol. 113, no.~3, pp.
  554--559, 2016.

\bibitem{amelkin2017polar}
V.~Amelkin, F.~Bullo, and A.~K. Singh, ``Polar opinion dynamics in social
  networks,'' \emph{IEEE Transactions on Automatic Control}, vol.~62, no.~11,
  pp. 5650--5665, 2017.

\bibitem{liu2018discrete}
J.~Liu, M.~Ye, B.~D. Anderson, T.~Basar, and A.~Nedic, ``Discrete-time polar
  opinion dynamics with heterogeneous individuals,'' in \emph{2018 IEEE
  Conference on Decision and Control (CDC)}.\hskip 1em plus 0.5em minus
  0.4em\relax IEEE, 2018, pp. 1694--1699.

\bibitem{spielman2009spectral}
D.~Spielman, ``Spectral graph theory ({Lecture 3: Laplacian and Adjacency
  Matrices}),'' \emph{Lecture Notes, Yale University: 740--0776}, 2009, \url
  {https://www.cs.yale.edu/homes/spielman/561/2009/lect03-09.pdf}, accessed
  2020-08-01.

\bibitem{meyer2000matrix}
C.~D. Meyer, \emph{Matrix analysis and applied linear algebra}.\hskip 1em plus
  0.5em minus 0.4em\relax Siam, 2000, vol.~71.

\bibitem{newman2018networks}
M.~Newman, \emph{Networks}.\hskip 1em plus 0.5em minus 0.4em\relax Oxford
  university press, 2018.

\bibitem{exex}
A.~Ozdaglar, ``Strategic form games and {N}ash equilibrium,''
  \emph{Encyclopedia of Systems and Control, Editors John Baillieul, Tariq
  Samad, Springer}, 2013.

\bibitem{krackhardt1987cognitive}
D.~Krackhardt, ``Cognitive social structures,'' \emph{Social networks}, vol.~9,
  no.~2, pp. 109--134, 1987.

\bibitem{horn1985c}
R.~A. Horn and C.~R. Johnson, \emph{Matrix Analysis}.\hskip 1em plus 0.5em
  minus 0.4em\relax Cambridge University Press, 2012.

\bibitem{moulin1984dominance}
H.~Moulin, ``Dominance solvability and cournot stability,'' \emph{Mathematical
  social sciences}, vol.~7, no.~1, pp. 83--102, 1984.

\bibitem{rosen1965existence}
J.~B. Rosen, ``Existence and uniqueness of equilibrium points for concave
  n-person games,'' \emph{Econometrica: Journal of the Econometric Society},
  pp. 520--534, 1965.

\end{thebibliography}
\end{document}